\documentclass[12pt,preprint]{aastex}
\usepackage{graphicx}
\usepackage{psfig}
\usepackage{natbib}

\newcommand{\gsim}{\mbox{\hspace{.2em}\raisebox{.5ex}{$>$}\hspace{-.8em}\raisebox{-.5ex}{$\sim$}\hspace{.2em}}}

\newcommand{\E}[1]{\times 10^{#1}}
\newcommand{\twCO}{$^{12}$CO}  \newcommand{\thCO}{$^{13}$CO}
\newcommand{\HII}{\mbox{H\,\textsc{ii}}}

    \newcommand{\Msun}{M_{\odot}}

     \newcommand{\du}{d_{4.4}}
\newcommand{\VLSR}{V_{\rm LSR}}

\begin{document}

\title{
The Dense Filamentary Giant Molecular Cloud G23.0-0.4: Birthplace of Ongoing Massive Star
Formation
}

\shorttitle{GMC G23.0$-$0.4}

\author{
Yang Su\altaffilmark{1,2}, Shaobo Zhang\altaffilmark{1,2}, 
Xiangjun Shao\altaffilmark{1,3}, and Ji Yang\altaffilmark{1,2} 
       }

\affil{
$^1$ Purple Mountain Observatory, Chinese Academy of
Sciences, Nanjing 210008, China \\
$^2$ Key Laboratory of Radio Astronomy, Chinese Academy of
Sciences, Nanjing 210008, China \\
$^3$ Graduate University of the Chinese Academy of Sciences, 19A Yuquan Road, 
Shijingshan District, Beijing 100049, China \\
      }

\begin{abstract}
We present observations of 1.5 square degree maps of the \twCO,
\thCO, and C$^{18}$O~($J$=1--0) emission toward the complex region 
of the supernova remnant (SNR) W41 and SNR G22.7-0.2. A massive 
($\sim5\times10^{5}~\Msun$), large ($\sim84\times$15~pc), and dense
($\sim10^{3}$~cm$^{-3}$) giant molecular
cloud (GMC), G23.0$-$0.4 with $\VLSR\sim$77~km~s$^{-1}$, is found to be
adjacent to the two SNRs. The GMC displays a filamentary
structure approximately along the Galactic plane. The filamentary structure
of the dense molecular gas, traced by C$^{18}$O~($J$=1--0) emission, is also 
coincident well with the distribution of the dust-continuum emission in the direction. 
Two dense massive MC clumps, two 6.7~GHz methanol masers, and one 
\HII/SNR complex, associated with the 77~km~s$^{-1}$ GMC G23.0$-$0.4, 
are aligned along the filamentary structure, indicating the star forming activity 
within the GMC. These sources have periodic projected spacing of 0\fdg18--0\fdg26 along
the giant filament, which is consistent well with the theoretical predictions
of 0\fdg22. It indicates that the turbulence seems to dominate the
fragmentation process of the dense gaseous filament on large scale.
The established 4.4~kpc distance of the GMC and the long dense 
filament traced by C$^{18}$O emission, together with the rich massive star 
formation groups in the nearby region, suggest that G23.0$-$0.4 is 
probably located at the near side of the Scutum-Centaurus arm 
in the first quadrant. 
Considering the large scale and the elongation structure along the Galactic plane, 
we speculate that the dense filamentary GMC has relation to the spiral density
wave of the Milky Way.

\end{abstract}


\section{INTRODUCTION}
The molecular gas is mostly within the Galactic plane and the giant molecular 
clouds (GMCs) are more concentrated toward spiral arms of the Milky Way
\citep{1985ApJ...292L..19S,2006ApJ...641L.113S}. 
The GMCs, as tracers of the large-scale 
structure in the Galaxy, are birthplaces of most of massive stars 
in the Galactic plane. Although they are rare in the 
Galaxy, massive stars significantly affect their surroundings
\citep[e.g.,][]{2014ASSP...36..173W}. Determining the nature of massive star 
formation is difficult because of its complexities (e.g., the short 
lifetime, the obscuration by their parent dust clouds, and the many processes 
therein).
Therefore, understanding the properties of the GMCs is a key step 
toward understanding massive star formation.
Moreover, the dense GMCs are also helpful in constructing the large-scale
Galactic structure since they are a good tracer of the spiral structure
in the Milky Way \citep[e.g.,][]{2014ApJ...797...53G}.

\cite{2014ASSP...36..225A} recently synthesized a comprehensive physical 
picture to describe star formation in dense cores of filamentary networks 
of molecular clouds (MCs). Several cases were also studied in observations, 
e.g., the infrared dark cloud (IRDC) Nessie \citep{2010ApJ...719L.185J}, 
the Rosette GMC \citep{2012A&A...540L..11S}, 
the IRDC G14.225$-$0.506 \citep{2013ApJ...764L..26B}, 
the Cygnus OB 7 GMC \citep{2014ApJ...797...58D}, and the IRDC G011.11$-$0.12 
\citep{2015A&A...573A.119R}. 
\cite{2013ApJ...763...57T} summarized the hierarchical fragmentation
structure of the Orion Molecular Cloud (OMC) filaments from 
GMCs scale ($\approx$35~pc) to small-scale clumps scale ($\approx$0.3~pc).
Along the dense filament, massive stars probably form from these 
dense fragmentational regions \citep{2010ApJ...719L.185J}.
However, the fragmentation structures
of other long dense GMC filaments are less studied.
It is partly due to the complicated morphology of MCs, the superposition of
the emission of molecular gas along the line of sight (LOS), and the 
strong disruption of GMCs by the massive stars feedback within it 
\citep{2013A&A...559A..34L,2014A&A...568A..73R}.

On the other hand, it is powerful to combine the dust emission and the molecular
line dataset in investigating the nature of the filamentary structure.
In the paper, the column density distributions of the densest structures can benefit
from the ATLASGAL dust-continuum survey, while the kinematical information
about the dense filament can be gotten from our CO and their isotopes survey
(see Section 2).

GMC G23.0$-$0.4, which is centered at ($l$, $b$, $v$)=(23\fdg0, $-0$\fdg4, 77~km~s$^{-1}$) 
and roughly aligned with the Galactic Plane, has a 
coherent velocity structure over about 1\fdg1$\times$0\fdg2 as traced by 
\thCO~($J$=1--0) and C$^{18}$O~($J$=1--0) emissions 
\citep[Figure~2, also see][]{2014ApJ...796..122S}. 
Several \HII\ regions
\citep[see the \HII\ region catalogs in][]{1989ApJS...71..469L,
1996ApJ...472..173L,2011ApJS..194...32A,2014ApJS..212....1A}, 
a group of early-type massive stars\citep{2014A&A...569A..20M}, and two SNRs
\citep{2014BASI...42...47G} 
are located in the field of view (FOV) of the GMC. The distances of most of 
above objects are consistent with that of GMC G23.0$-$0.4  
\citep{2010ApJ...708.1241M,2014A&A...569A..20M,2014ApJ...796..122S}. 
G23.0$-$0.4, as a long dense GMC, is a good laboratory for the 
investigation of the properties of the filament.
The locally massive stars feedback near/within the dense GMC also give
us a good opportunity to study the relationship between them.

In the paper, we mainly
focus on the nature of the filamentary GMC G23.0$-$0.4 in the complicated 
interstellar medium (ISM).
We also discuss the relationship between the GMC and the 
ambient star forming activity.
We present the observations of \twCO~($J$=1--0), \thCO~($J$=1--0), and 
C$^{18}$O~($J$=1--0) and the data reduction in the following
section. Section 3 shows the properties of the GMC. In Section 4, we mainly 
discuss the fragmentation process along the dense filament. 
We finally summarize our conclusions in Section 5.

\section{OBSERVATIONS AND DATA REDUCTION}

Observations in the \twCO~($J$=1--0), \thCO~($J$=1--0), and 
C$^{18}$O~($J$=1--0) were made simultaneously with the 13.7~m 
millimeter-wavelength telescope, located at Delingha in China.
The 9-beam Superconducting Spectroscopic Array Receiver (SSAR)
was working as the front end in sideband separation mode
\citep[see the details in][]{shan}. 

The data were observed using the
on-the-fly (OTF) mode, with the standard chopper wheel method for 
calibration. In the mode, the telescope beam is scanned along 
lines in the Galactic longitude and latitude directions on
the sky at a constant rate of 50$''$~s$^{-1}$ and the receiver 
records spectra every 0.3 sec. The data scanned in both 
longitude and latitude directions were combined together
to reduce the fluctuation of noise perpendicular to the scanning 
direction. 

We adopted the main beam efficiency $\eta_{\rm mb}$=0.44 for
\twCO~($J$=1--0) and 0.48 for \thCO~($J$=1--0) and C$^{18}$O~($J$=1--0). 
The mean rms noise
level of the calibrated brightness temperature ($T_{\rm R}$) was 
$\sim$ 0.5~K for \twCO~($J$=1--0) at a resolution of 0.16~km~s$^{-1}$ and 
$\sim$ 0.3~K for \thCO~($J$=1--0) and C$^{18}$O~($J$=1--0) at 0.17~km~s$^{-1}$. 
All CO data were reduced using the GILDAS/CLASS
package\footnote{http://www.iram.fr/IRAMFR/GILDAS}.

The available CO High Resolution Survey \citep[COHRS,][]{2013ApJS..209....8D} 
data 
and the Galactic Ring Survey \citep[GRS,][]{2006ApJS..163..145J} data were also 
used for comparison. The 1.4~GHz VLA Galactic Plane Survey 
\citep[VGPS,][]{2006AJ....132.1158S} data and the 870~$\mu$m 
ATLASGAL survey \citep{2009A&A...504..415S} data were used to trace the 
thermal/non-thermal radio emission and the cold dust emission, respectively.
We also use the \mbox{H\,\textsc{i}} data of VGPS to resolve the 
kinematic distance ambiguity of GMC G23.0$-$0.4 (see Section 4.1). 
We briefly summarized the information of these survey data in Table~1.

\section{RESULTS}
The molecular gas toward the observational region contains a number
of velocity components, which indicates the overlapping MCs along the LOS
in the complicated region. It is unsurprising when we consider that 
a large amount of molecular gas is located in the inner Galaxy. 
In the direction of $l$=23\fdg0, our LOS
will across several spiral arms (the local/Orion-Cygnus arm, the
Carina-Sagittarius arm, the Scutum-Centaurus arm, and the Perseus arm) and
the Galactic Bar \citep[e.g.,][]{2000MNRAS.317L..45H,2005ApJ...630L.149B}.

In Figure~1, we show the typical spectra, which is extracted from a
30$'\times30'$ region centered at ($l$=23\fdg25, $b$=$-$0\fdg25).
We find that, even for optically thin C$^{18}$O~($J$=1--0) line, the
spectrum is crowded in the velocity interval of 50--110~km~s$^{-1}$.
Four distinct C$^{18}$O peaks can be discerned at 52.6, 63.6, 77.4,
and 96.8~km~s$^{-1}$, which probably indicate the different MC components along 
the LOS. Also, several other MC components (e.g., 57.4, 81.6, and 101.2~km~s$^{-1}$) 
can be seen near the above C$^{18}$O peaks.

The complicated MC components in the velocity interval of 50--110~km~s$^{-1}$
prevent us to further discuss the relationship between the different MCs.
Fortunately, in the velocity range of 70--80~km~s$^{-1}$, the C$^{18}$O~($J$=1--0)
emission of GMC G23.0-0.4 that we are interested in suffer less from contamination 
from other MC components along the LOS. The filamentary structure of the GMC 
can be confirmed to be coherent in velocity space, especially from the 
C$^{18}$O emission (Section 3.1). Moreover, we have shown that the distance 
of the GMC with LSR velocity of 77~km~s$^{-1}$ was determined (see Section 4.1).
The established distance allows us to investigate more reliable
physical properties of the GMC accordingly.
In the section, we mainly study the properties of GMC G23.0$-$0.4 
($\VLSR\sim$70--84~km~s$^{-1}$; Section 3.1).
Other interesting MCs adjacent to the filamentary GMC, which have similar
LSR velocity of $\sim$77~km~s$^{-1}$, are also
investigated (Section 3.2).

\subsection{Dense Filamentary GMC G23.0$-$0.4}
\subsubsection{Distribution of the CO Emission}
We made the 3-color intensity image in the velocity interval of 
$\VLSR$=74--79~km~s$^{-1}$, overlaid on the VGPS 1.4~GHz
radio continuum contours (Figure~2). 
It is clearly seen that the overall distribution of the molecular gas is
quite complicated in the FOV (Figure~2).
We also note that the characteristics
of our \thCO~($J$=1--0) emission are consistent well with that of GRS data.
Meanwhile, the C$^{18}$O~($J$=1--0) emission in our observations can provide
further insight into the nature of the dense molecular gas.
Here we mainly focus on the brightest filamentary structure: GMC G23.0$-$0.4.

In the 3-color image, the
\twCO\ ($J$=1--0) emission in blue is more diffuse and extended than the 
\thCO~($J$=1--0; green) and C$^{18}$O~($J$=1--0; red) emission. 
In contrast, the C$^{18}$O~($J$=1--0) 
emission is only seen in the region of bright 
\thCO~($J$=1--0) emission. It is probably because the typical densities traced
by them are different \citep[e.g., $\sim10^{2}$~cm$^{-3}$ for optically
thick \twCO~($J$=1--0), $\sim10^{3}$~cm$^{-3}$ for median optical depth 
\thCO~($J$=1--0), and $\sim10^{4}$~cm$^{-3}$ for optically thin 
C$^{18}$O~($J$=1--0); e.g.,][]{2005ApJ...634..476Y}. We thus expect that the 
\twCO~($J$=1--0), \thCO~($J$=1--0), and C$^{18}$O~($J$=1--0) emission is 
from the enveloping layer of low density, the middle layer of intermediate 
density, and the inner core of high density of a MC, respectively. 
The C$^{18}$O~($J$=1--0) emission, which is a good tracer of the 
high-density molecular gas ($\sim10^{4}$~cm$^{-3}$), can reveal the nature of the 
dense part of a GMC. The blue envelopes, the green intermediate layers, 
and the red dense cores of the MCs are indeed distinguished in Figure~2. Therefore, 
it is useful in studying the properties of the GMC by combining the 
\twCO~($J$=1--0), \thCO~($J$=1--0), and C$^{18}$O~($J$=1--0) dataset.

The brightest portion
of GMC G23.0$-$0.4, which appears roughly parallel to the Galactic plane,
generally displays a long filamentary structure. The dense gas of the GMC 
traced by C$^{18}$O~($J$=1--0) emission is primarily distributed in a rectangle region 
centered at ($l$, $b$, $v$)=(23\fdg0, $-0$\fdg4, 77~km~s$^{-1}$) (see the red rectangle 
in Figure~2). Actually, the main axis of GMC G23.0$-$0.4 is inclined by $20^{\circ}$ 
with respect to the Galactic plane. The dense filament of GMC G23.0$-$0.4 extends 
$\sim$1\fdg1 in length and is distinguished by bright CO emissions. 
The mean width of the filamentary GMC traced by C$^{18}$O~($J$=1--0) emission
is about 6$'$, whereas the width of the thin filament in the denser region
is near 2$'$ or less. It indicates that there are some substructures (e.g., slim
filaments, dense cores and clumps) within the filamentary GMC G23.0$-$0.4 (Figure~2).
Future observations at higher spatial resolution will be helpful in investigating
these interesting characteristics of the GMC.

In general, the filamentary structure of GMC G23.0$-$0.4 is hierarchical.
In our spatial resolution ($\sim$0\farcm8), GMC G23.0$-$0.4 displays a
wave-like and branch structures along the trunk of the filament. Some thin
filaments with $\sim2'$ width and pc-scale dense clumps are distributed
along the 77~km~s$^{-1}$ giant filament (e.g., see Figures 2, 3, and 4).

\subsubsection{Physical Parameters of the GMC}
Physical parameters of the GMC can be derived from the following
description \citep[e.g., see the appendix in][]{1997ApJ...476..781B}.
Firstly, the excitation temperature of the CO gas $T_{\rm ex}$ can be derived
using the equation: \\

$T_{\rm mb} = f [J(T_{\rm ex}) -J(T_{\rm bg})] [1-{\rm exp}(-\tau)]$, \\

where $\tau$ is the optical depth, $T_{\rm ex}$ and $T_{\rm bg}$ is the excitation
temperature and the background temperature, respectively, and \\

$J(T) = \frac{h\nu}{k} \frac{1}{{\rm exp}(h\nu/kT) - 1} $.\\

Assuming that the \twCO~($J$=1--0) emission is optically thick ($\tau_{12}\gg1$)
and the beam filling factor $f$ of it is equal to 1, we can 
get the excitation temperature from the peak temperature of \twCO~($J$=1--0)
emission.
The distribution of the excitation temperature is shown in Figure~3.
The self-absorption from the cold foreground gas along the LOS is ignored and 
the derived excitation temperature as a result should be regarded as lower limit.  

The optical depth of \thCO~($J$=1--0) and C$^{18}$O~($J$=1--0) of GMC
G23.0$-$0.4 can be estimated by assuming that the excitation 
temperature of these molecular lines have the same value as that of 
\twCO\ pixel by pixel (see Figure~3). Thus, in the 
local thermodynamic equilibrium (LTE) assumption, the optical
depths of $\tau_{13}$ and $\tau_{18}$, the column densities of 
\thCO\ and C$^{18}$O can be derived using the
following equations \citep[e.g., see the appendix in][]{1997ApJ...476..781B}: \\

$\tau_{13}(V) = - {\rm ln}[1 -\frac{T_{\rm MB13}(V)}{5.29[J_{13}(T_{\rm ex})-0.164]}]$, \\

$\tau_{18}(V) = - {\rm ln}[1 -\frac{T_{\rm MB18}(V)}{5.27[J_{18}(T_{\rm ex})-0.167]}]$, \\

$N({\rm ^{13}CO}) = 2.42 \times 10^{14} \sum\limits_{V} [ \frac{0.2({\rm km~s^{-1}}) \tau_{13}(V) T_{\rm ex}}{1-{\rm exp}[-5.29/T_{\rm ex}]} ]~{\rm cm}^{-2} $, \\

and\\

$N({\rm C^{18}O}) = 2.42 \times 10^{14} \sum\limits_{V} [ \frac{0.2({\rm km~s^{-1}}) \tau_{18}(V)  T_{\rm ex}}{1-{\rm exp}[-5.27/T_{\rm ex}]} ]~{\rm cm}^{-2} $, \\

where $T_{\rm ex}$ is the excitation temperature of these molecules in K, and
$J_{13}(T_{\rm ex})=1/[{\rm exp}(5.29/T_{\rm ex}) -1]$ for \thCO~($J$=1--0) and 
$J_{18}(T_{\rm ex})=1/[{\rm exp}(5.27/T_{\rm ex}) -1]$ for C$^{18}$O~($J$=1--0).
In the calculation, we divide each spectrum into 0.2~km~s$^{-1}$ bins 
to estimate the optical depth in each bin. 
Accordingly, we can calculate the \thCO\ and C$^{18}$O column
density within the LSR velocity range from 72 to 81~km~s$^{-1}$.

The H$_2$ column density of the GMC can be estimated from
the optically thin \thCO\ ($J$=1--0) and C$^{18}$O~($J$=1--0) emission.
In the above calculation, the \thCO\ abundance of
$N$(H$_2$)/$N(^{13}$CO) $\approx7\E{5}$ \citep{1982ApJ...262..590F}
and the C$^{18}$O abundance of
$N$(H$_2$)/$N$(C$^{18}$O) $\approx7\E{6}$ \citep{1995A&A...294..835C}
are adopted.
On the other hand, using the optically thick \twCO\ ($J$=1--0) emission as the tracer of
molecular gas, the H$_2$ column density of the GMC can be
estimated directly by adopting the mean CO-to-H$_2$ mass conversion factor
$1.8\E{20}$~cm$^{-2}$K$^{-1}$km$^{-1}$s \citep{2001ApJ...547..792D}.
In the estimation of the mass of the GMC, a mean molecular weight per
H$_2$ molecule of 2.76 has been adopted. The distance of the GMCs
was adopted as 4.4~kpc (see Section 4.1).
The parameters of the filamentary GMC G23.0$-$0.4 are listed in Table~2.

According to Figure~3, we find that the excitation temperature in the region of 
the dense filamentary GMC G23.0$-$0.4 is higher compared to that of the other 
molecular gas in the FOV. The mean excitation temperature of the filamentary structure 
is about 20~K and the excitation temperatures in some interface regions between SNR W41
and SNR G22.7$-$0.2 are higher than 26~K.
The value of the excitation temperature of GMC G23.0$-$0.4 is comparable
to those of other GMCs, e.g., 10--70~K for the Orion-A GMC \citep{1998AJ....116..336N}
and 7--56~K for the W51 GMC \citep{2012MNRAS.424.1658P}. We also note that
the excitation temperature of GMC G23.0$-$0.4 is very close to 
that in the high density layer of the W3 GMC \citep[15--30~K,][]{2012MNRAS.422.2992P}.

The structures of
GMC G23.0$-$0.4 can be distinguished readily in the maps
of the optical depth of the C$^{18}$O~($J$=1--0) emission (the right panel 
in Figure~4). We find that the distribution of $\tau_{13}$ is 
more diffuse and extended than that of $\tau_{18}$. $\tau_{13}$ in the dense part of 
GMC G23.0$-$0.4 is larger than 0.5, which indicates that the optical depth of 
\thCO~($J$=1--0) is not optically thin. On the other hand, $\tau_{18}$ is 
mostly less 0.15 in the region of the filamentary GMC G23.0$-$0.4. The 
highest $\tau_{18}$ are 0.21--0.23 in the high $\tau_{13}$ region, which 
show the densest parts in the GMC.
The distribution of the C$^{18}$O gas is more compact than those of the 
\twCO\ and \thCO\ gas, which indicates that the C$^{18}$O emission
can clearly reveal dense structures of GMC G23.0$-$0.4 indeed.

The column density, the mass, and the density of the main body of
GMC G23.0$-$0.4 is 1.7--2.9$\times10^{22}$~cm$^{-2}$,
3.2--5.2$\times10^{5}~\Msun$, and 730--1200~cm$^{-3}$, respectively.
We note that the column density only represents the mean value for the
filamentary GMC in the rectangle region. The column density is higher
than 5$\times10^{22}$~cm$^{-2}$ in some dense regions, which indicates that
the volume density of these regions is higher than 6$\times10^{3}$~cm$^{-3}$
(assuming a depth of 2$'$, the mean width of slim filaments, for the dense
region). Therefore, the strong C$^{18}$O~($J$=1--0) emission,
which represents the distribution of the high-density molecular gas in the
GMC G23.0$-$0.4, is very likely tracing the potential star formation
regions (see Section 4.3.1).

The mass of the GMC listed in Table~2 is probably the lower limit
because some diffuse molecular gas outside the rectangle region
is not accounted for. 
Based on the optically thick \twCO\ ($J$=1--0) emission, we estimate that 
the total mass of GMC G23.0$-$0.4 is $\sim1.2\times10^{6}\Msun$ due to 
a large amount of low-density molecular
gas in the enveloping layer of the GMC, which is outside the rectangle region.
In the above calculation, all pixels above 40~K~km~s$^{-1}$ (about the half
of the \twCO\ mean intensity of the GMC) are accounted for
and the total mass of the GMC is derived from a
$\sim$2200~arcmin$^2$ region, in which region the velocity interval is 
68--82~km~s$^{-1}$ and the mean intensity is 87~K~km~s$^{-1}$, respectively.
On the contrary, the dense gas (e.g., $\gsim10^{3}$--$10^{4}$~cm$^{-3}$), which
only represents a fraction of the GMC's total mass, is mainly located in the
trunk of the filamentary GMC G23.0$-$0.4 (388~arcmin$^2$ in Table~2,
also see the right panel of Figure~4). 

The filamentary GMC G23.0$-$0.4 is
different from the giant molecular filaments studied by \cite{2014A&A...568A..73R},
in which samples these giant filaments have low dense gas mass fractions (2\%--12\%).
In the analysis of \cite{2014A&A...568A..73R}, they selected the \thCO\ ($J$=1--0) 
emission ($<10^{3}$~cm$^{-3}$) as the total masses tracer and the ATLASGAL
870~$\mu$m dust emission as the dense gas mass tracer ($>10^{22}$~cm$^{-2}$). 
In our study, we use \twCO\ ($J$=1--0) emission ($\sim10^{2}$~cm$^{-3}$) and
C$^{18}$O~($J$=1--0) emission ($>10^{22}$~cm$^{-2}$, see Figure~9) to calculate 
the total mass and the dense gas mass of the GMC, respectively. The total mass of 
the GMC in our study is obviously larger than that from the \thCO\ ($J$=1--0) emission,
while the mass of the dense gas from the C$^{18}$O~($J$=1--0) emission is 
roughly compared to that from the ATLASGAL 870~$\mu$m dust emission 
(see the discussion of FP1 in Section 4.3.1). 
Therefore, GMC G23.0$-$0.4 seems to be denser than the GMC samples studied
by \cite{2014A&A...568A..73R}.

\subsubsection{Velocity Structures of the Dense Molecular Gas}
We made the channel maps of C$^{18}$O~($J$=1--0) emission in the interval of 
69--85~km~s$^{-1}$ by step of 2~km~s$^{-1}$ to investigate the spatial 
distribution of the dense 
molecular gas (Figure~5). The main body of filamentary GMC G23.0$-$0.4 is in 
the velocity interval of 75--79~km~s$^{-1}$. The molecular gas in 71--75~km~s$^{-1}$ 
is assembled in the interface between SNR W41 and SNR G22.7$-$0.2. On the other
hand, the molecular gas in 79--85~km~s$^{-1}$ are roughly in the centre
of SNR W41. 

In the position--velocity (PV) diagrams along the long filamentary GMC 
G23.0$-$0.4 (Figure~6), we can discern that the GMC is in the velocity interval of 
70--84~km~s$^{-1}$. The interaction between SNR G22.7$-$0.2 and the GMC can be 
discerned from the \twCO~($J$=1--0) PV diagram \citep[see detail in][]{2014ApJ...796..122S}. 
The $\VLSR\sim$100~km~s$^{-1}$ 
MC component is from the molecular gas near the tangent point
\citep[MC G23.4,][]{2012PASJ...64...74O}, which is irrelevant to GMC G23.0$-$0.4.
We also made the PV diagram of the C$^{18}$O~($J$=1--0) emission along the interface 
between SNR W41 and SNR G22.7$-$0.2 (Figure~7). The PV diagram
delineates some diffuse emission extended to 70~km~s$^{-1}$ besides the dense
74 and 78~km~s$^{-1}$ MCs. The feature probably relates to 
the star forming activity near the region.
Based on the spatial distribution of the molecular gas, it seems reasonable 
to assume that the emission within the velocity of 70--84~km~s$^{-1}$ comes 
from GMC G23.0$-$0.4 that is associated with the ambient SNRs \citep{2014ApJ...796..122S}.
The distance estimation of the MCs suggests that the GMC complex 
is probably in the near side of the Scutum-Centaurus arm (see Section 4.1).

\subsection{Other Interesting MCs}
Besides GMCs G23.0$-$0.4 (this paper) and G22.6$-$0.2 \citep{2014ApJ...796..122S}, 
there are some interesting MCs with similar LSR velocity in the FOV.
A partial shell structure ($l$=23\fdg4, $b$=$-0$\fdg6) is ambient to 
the southern edge of SNR W41, and a series of pillar-like protrusions are along 
a bright-rim structure near ($l$=23\fdg25, $b$=0\fdg08) (Figure~2).

The low-density ($\sim10^{2}$--$10^{3}$~cm$^{-3}$) molecular gas 
of the partial shell structure is in the southern boundary of SNR W41, in which
place the remnant shows very faint radio emission
(see the left panel in Figure~4).
An early-type massive star would create a wind bubble of radius 
of $56n_{\rm ISM}^{-0.3}$~pc \citep{1984ApJ...278L.115M}.
A radius of such a bubble will be about 23~pc,
assuming the mean density of the region is 20~cm$^{-3}$.
This value is less than that of the partial shell structure
(radius of $\sim28'$ or 36~pc at a distance of 4.4~kpc). 
Therefore, the partial shell structure, with a total mass of
$\sim2\times10^{5}~\Msun$, is likely the result of a series of massive 
stellar winds. On the other hand, supernova explosions are also probably held
responsible for the formation of such structure.   

In the northern region of the FOV, some small pillar-like protrusions
seem to be located along a partial shell near ($l$=23\fdg25, $b$=0\fdg08).
These small pillar-like protrusions have bright \thCO~($J$=1--0) and
C$^{18}$O~($J$=1--0) emission (see Figure~4), which can be distinguished
from the diffuse \twCO~($J$=1--0) surroundings (Figure~2). 
At the apexes of the pillar-like molecular clouds, bright
870~$\mu$m dust emission (5.94~Jy for G23.1989$+$0.0009 and 4.85~Jy
for G23.2646$+$0.0774) are coincident well with the peak emission
of C$^{18}$O~($J$=1--0). The interesting structures
are probably related to the energetic radiation of the nearby 
early-type stars. Such bright-rim clouds are often associated
with the photoevaporation from the nearby massive star 
\citep[e.g.,][]{1999ApJ...513..339T,2010A&A...518L..90B}.
A diffuse \HII\ region G23.162$+$0.023 \citep{1996ApJ...472..173L},
which is indeed located at the west of these pillar-like MCs,
is probably responsible for these bright-rim clouds.

Both of the interesting MCs adjacent to GMC G23.0$-$0.4 are 
very likely associated with stellar feedback from massive stars. 
It shows that GMC 23.0$-$0.4 is sited near the region of massive stars
in the view of the adjacent interesting MCs, which is consistent with 
the result of Section 4.2.

\section{DISCUSSION}
\subsection{Distance of GMC G23.0$-$0.4}

We use two methods to determine the distance of the dense filamentary GMC G23.0$-$0.4.
In the first method, the distance of the GMC can be derived from the
trigonometric parallax assuming the association between the GMC
and the 6.7~GHz methanol maser G23.01$-$0.41. The distance of it
is $4.59^{+0.38}_{-0.33}$~kpc \citep{2009ApJ...693..424B}. 
It appears reasonable because the 6.7~GHz methanol 
maser G23.01$-$0.41, which is just located in the centre of the GMC,
has a similar LSR velocity 
{\citep[V$_{\rm maser}\sim$81~km~s$^{-1}$, see Table 2 in][]{2009ApJ...693..424B} 
to GMC G23.0$-$0.4.  

In the second method, the distance of the GMC can be estimated from
the LSR velocity ($\sim$77~km~s$^{-1}$) of it. Using the Galactic rotation curve
of \cite{2014ApJ...783..130R}, we thus place the GMC
G23.0$-$0.4 at a near kinematic distance of 4.4$\pm$0.4~kpc.
The kinematic distance ambiguity can be resolved from the \mbox{H\,\textsc{i}}
self-absorption method \citep[HISA, see Figures 1 and 2 in][]{2009ApJ...699.1153R}. 
Based on the HISA method, 
a molecular cloud located at the near kinematic distance
will exhibit the 21~cm absorption feature that is coincident
with the \thCO\ peak from the cloud \citep{2009ApJ...699.1153R}. These authors 
suggested that the $\sim70-82$~km~s$^{-1}$ MCs 
(see Table~1 in their paper), which is mostly associated with GMC G23.0$-$0.4
in the FOV, is located at the near kinematic distance. We also 
checked the method for other small 
regions in GMC G23.0$-$0.4 using VGPS \mbox{H\,\textsc{i}} data and 
our CO data and find that the GMC is indeed in the near side. 
Accordingly, we exclude the far kinematic distance of the GMC, which result is 
consistent with that of \cite{2013ApJ...770...39E}. 
In their work, the $\sim$77~km~s$^{-1}$ GMC is 
also in the near side \citep[see Table~3 in][]{2013ApJ...770...39E}.

We note that the near kinematic distance of the GMC is in agreement well
with the trigonometric distance and that of other works
\citep{2008AJ....135..167L,2010ApJ...708.1241M,2014A&A...569A..20M}. 
Combining the above analysis, we suggest that the dense filamentary 
GMC G23.0$-$0.4, together with the rich massive star formation groups ambient to
the giant filament (Section 4.2), is probably located at the 
near side of the Scutum-Centaurus arm in the first quadrant 
\citep[e.g., please refer to the spiral arm models of the Milky Way,][]
{1993ApJ...411..674T,2014ApJ...783..130R}. 
For simplicity, we adopt the value of 4.4~kpc as the distance 
of the GMC throughout the paper.

\subsection{Environment of GMC G23.0$-$0.4}
We have shown that SNRs G22.7$-0.2$ and W41 are both interacting with 
the $\VLSR\sim$77~km~s$^{-1}$ GMC G23.0$-$0.4 \citep{2014ApJ...796..122S}. 
The kinematic signature of the interaction between SNR G22.7$-0.2$ and
the GMC can be seen from the PV diagram of the \twCO~($J$=1--0) emission
(the left panel of Figure~6). On the other hand, emission of \twCO~($J$=3--2) 
is a good tracer of the warm and dense
molecular gas associated with star formation. It is also a good tracer
of shocked gas, e.g., outflow activity, SNR--MC interaction, and
cloud--cloud collision. Unfortunately, the COHRS
version 1 only maps a region of $|b|<$0\fdg25 in the direction \citep{2013ApJS..209....8D},
which does not cover the main body of GMC G23.0$-$0.4. Nevertheless,
the available \twCO~($J$=3--2) data are used to study the molecular gas in the FOV.
Actually, some enhanced \twCO~($J$=3--2) emission in the northern region of
SNR W41 is detected in the velocity interval 70--74~km~s$^{-1}$ (the blueshift
compared to the LSR velocity 77~km~s$^{-1}$ of the GMC)
when we searched for the proof of the SNR-MC interaction. It is
consistent with the result that SNR W41 lies behind GMC G23.0$-$0.4 and
the remnant is interacting with the GMC \citep{2013ApJ...773L..19F}.
We hope that the data of the COHRS release 2 will provide further insight into
the nature of the warm and dense molecular gas of the overall GMC G23.0$-$0.4.

There are multiple overlapping \HII\ regions in the FOV of 
SNRs G22.7$-$0.2 and W41 \citep[e.g.,][]{2010ApJ...708.1241M}. 
Even though the velocity of \HII\ regions often deviates from that of
their parent MCs, we suggest that most of them with the 
systematic velocity of 70--81~km~s$^{-1}$ probably have similar distance to
the 77~km~s$^{-1}$ GMC \citep{2014ApJ...796..122S}. 
In the right panel of Figure~8, positions of these regions of twelve 
\HII\ regions and two \HII/SNR complexes \citep{1989ApJS...71..469L,1997ApJ...488..224K,
2005AJ....129..348G,2005AJ....130..156G,2006AJ....131.2525H,2011ApJS..194...32A,
2014ApJS..212....1A} are marked with red and blue circles, respectively. 
Actually, at least 6 \HII\ regions \citep[e.g., see Table~6 in][]{2014ApJS..212....1A}
in the FOV are most likely located at the near distance.

Moreover, the stellar cluster GLIMPSE9 ($l$=22\fdg756, $b$=$-$0\fdg400) 
is also associated with GMC G23.0$-$0.4 \citep{2010ApJ...708.1241M,
2014ApJ...796..122S}. Recently, \cite{2014A&A...569A..20M} discovered a number 
of massive stars surrounding the cluster GLIMPSE9 in the south of SNR G22.7$-$0.2 
(REG GLIMPSE9Large in their paper) and some massive stars near the centre 
of SNR W41 (REG4 in their paper), which early-type stars with K-band extinction 
from $\sim$1.3--1.9 mag have spectrophotometric distances consistent with that of 
the GMC. The two massive star groups, which are most likely associated with the 
GMC G23.0$-$0.4 \citep{2010ApJ...708.1241M,2014A&A...569A..20M}, are also marked 
with the blue boxes in Figure~8. 

The existence of two SNRs, several \HII\ regions, and a number of massive stars 
associated with GMC G23.0$-$0.4 show the energetic star formation activity in 
the region. The natal GMCs of these objects are very likely destroyed by the 
process of the photoevaporation of the HII regions \citep{2007ARA&A..45..565M}. 
On the other hand, the supernova explosions will remove a significant
fraction of the cloud mass if the massive progenitors of them did not
run far away from their parent GMCs \citep{2015A&A...576A..95I}. 
However, these energetic process have a limited influence on the
nearby dense gas of the filamentary GMC G23.0$-$0.4. For example,
SNR G22.7$-$0.2, which's center is about 18~pc away from the dense filamentary GMC,
has only 3.3$\times$10$^{4}~\Msun$~km~s$^{-1}$ momentum and 
5$\times10^{48}$~erg kinetic energy injection to GMC G23.0$-$0.4
\citep{2014ApJ...796..122S}. The dense gas within GMC G23.0$-$0.4 still keep 
the filamentary structure, which is roughly parallel to the Galactic plane.

The formation of GMCs along the spiral arms is believed
to be related to the spiral shocks
\citep[e.g.,][]{2006MNRAS.371.1663D,2007MNRAS.376.1747D}. 
Assuming that GMC G23.0$-$0.4 and their ambient GMCs, which are the natal
places of the massive progenitors of the SNRs,
formed simultaneously in the region of the FOV, the age of GMC G23.0$-$0.4
is $\gsim$10~Myr (the typical lifetimes of B1--O9 stars), in which timescale the 
natal GMCs of SNRs G22.7$-$0.2 and W41 and the \HII\ regions \citep{2014ApJ...796..122S}
had been destroyed by the violent massive star formation therein 
\citep[e.g., Section 3.2.2 in][]{2007ARA&A..45..565M}.
But these energetic process outside GMC G23.0$-$0.4 have a limited influence on the
dense gas of the GMC.
On the other hand, the strong massive star forming
activities are ongoing along the filamentary GMC G23.0$-$0.4 (see Section 4.3),
which will exhaust and dissipate the molecular gas of the GMC in future several Myr.
Actually, a part of the dense gas has been
destroyed by the \HII/SNR complex G022.760$-$0.485 (FP2 in Figure~9, Section 4.3). 
In the region of the \HII/SNR complex G022.760$-$0.485, the C$^{18}$O~($J$=1--0)
emission is very weak \citep[see the region of 06 in Figure~8 of][]{2014ApJ...796..122S}, 
which is probably due to the stellar feedback of massive star therein. 
Accordingly, the lifetime of the dense filamentary GMC should be about 10--20~Myr,
which is consistent with the result of 1--3 free-fall time 
of massive GMCs \citep{2011ApJ...729..133M}.

\subsection{Fragmentation of the Giant Filament}
\subsubsection{Fragmentation on large scale}
Actually, C$^{18}$O~($J$=1--0) emission is not a good tracer for 
very high-density molecular gas ($\gsim10^{5}$~cm$^{-3}$) due to it's 
relatively low critical density. 
Other high-density tracers such as HCN or N$_{2}$H$^{+}$ are helpful in investigating
conditions within high-density regions of the GMC ($\sim10^{4}$--$10^{6}$~cm$^{-3}$). 
In our study, we use the bright dust emission to reveal the dense, massive structures
within GMC G23.0$-$0.4 accordingly. The strong dust emission is labelled on
red contours in the left panel of Figure~8. These dense 
molecular clouds, together with massive star formation regions,
probably represent the molecular gas concentration along the dense filament.

The fragmentation points (FPs) 1--5 along GMC G23.0$-$0.4 are shown
in Figure~9. The properties of these FPs are summarized in Table~3. 
FP1 contains a millimetric radio source BGPS G022.548$-$00.525
\citep{2010ApJS..188..123R},
in which region the emission from the 870~$\mu$m ATLASGAL data is also bright 
\citep[G022.5483$-$0.5225, 3.68~Jy;][]{2014A&A...565A..75C}.  
The mass of the dense molecular gas in the region of 1\farcm5$\times$1\farcm5
estimated from C$^{18}$O~($J$=1--0) emission is $\sim4\times10^{3}~\Msun$. 
Assuming that the dust can be characterized by a single temperature of 15~K 
in the region (see Figure~3), 
we calculate the isothermal mass of $\gsim6\times10^{2}~\Msun$
in the region of 0\farcm5$\times$0\farcm5 based on the dust-continuum data. 
The high surface density ($\gsim10^{3}\Msun$~pc$^{-2}$) 
implies the potential star formation in the region of FP1.
FP2 is a \HII/SNR complex region \citep{1989ApJS...71..469L,1997ApJ...488..224K,
2006AJ....131.2525H,2011ApJS..194...32A,2014ApJS..212....1A}, which is associated with the 
GMC G23.0$-$0.4 \citep{2010ApJ...708.1241M,2014A&A...569A..20M,2014ApJ...796..122S}. 
FP3 and FP4, contains the 6.7~GHz methanol masers G23.01$-$0.41 and G23.19$-$0.38,
respectively, which are the good tracers of massive star formation
\citep{2002A&A...392..277S,2009ApJ...693..424B}. In the two regions,
the 1.1~mm \citep{2010ApJS..188..123R} and 870~$\mu$m \citep{2014A&A...565A..75C} emission 
are very bright (e.g., BGPS G023.012$-$00.410 and ATLASGAL G023.0063$-$0.3991 (3.79~Jy),
and G023.0082$-$0.4092 (12.76~Jy) in FP3; BGPS G023.208$-$00.378 
and ATLASGAL G023.2056$-$0.3772 (11.30~Jy) and G023.2122$-$0.3726 (1.34~Jy) in FP4).  
FP5 is similar to that of FP1, in which place a millimetric radio source 
BGPS G023.368$-$00.290 is in the region \citep{2010ApJS..188..123R}. The emission 
from the 870~$\mu$m ATLASGAL data is also 
bright \citep[G023.3652$-$0.2887, 4.93~Jy;][]{2014A&A...565A..75C}
and many young stellar
objects (YSOs) are distributed in the vicinity of FP5 \citep{2009ApJ...698..324R}.

We also showed the abundance ratio distribution between \thCO\ and
C$^{18}$O, $X_{^{13}{\rm CO}}$/$X_{{\rm C}^{18}{\rm O}}$ for the
77~km~s$^{-1}$ molecular gas in the left panel of
Figure~8. We found that the abundance ratio
$X_{^{13}{\rm CO}}$/$X_{{\rm C}^{18}{\rm O}}$ becomes lower 
\citep[similar to the value of 5.5 in the Solar system, 
see Table~4 in][]{1994ARA&A..32..191W} in the regions of
the bright dust-continuum emission (the red contours in the left panel of Figure~8),
which probably indicates the high-density regions for star formation.
The difference between the abundance ratios in the bright dust emission regions
(e.g., $\sim$5 for FPs 1, 3, 4, and 5) and the other region 
(e.g., $\sim$10 for FP 2) very likely represents the evolutionary sequence of the
dense molecular gas in the GMC. The abundance ratio will be relatively higher 
when the dense molecular cores are chemically influenced by the far ultraviolet 
(FUV) radiation from the massive stars embedded in the GMC 
\citep{1988ApJ...334..771V}. In other words,
The photodissociation rate of the C$^{18}$O molecule is larger
than that for \thCO, which leads to the higher abundance ratio 
$X_{^{13}{\rm CO}}$/$X_{{\rm C}^{18}{\rm O}}$ in the massive star's
UV radiation field. 
It is consistent with the result that a part of the dense gas in
GMC G23.0$-$0.4 has been destroyed by the \HII/SNR complex G022.760$-$0.485 
(see Section 4.2). In contrast, the lower
abundance ratio represents the earlier stage of dense MCs, which are not or
less affected by massive star formation. The low abundance ratio in the
bright dust emission regions is also consistent with the result that most of
cold dust emission is indeed from 77~km~s$^{-1}$ GMCs (see Figure~8).
We find that the distribution of the dust clumps/cores (the fifteen 
black boxes in Figure~8) is coincident well with that of the C$^{18}$O emission.
In the FOV, 11/15 dust clumps are associated with GMC G23.0$-$0.4.
The distance of these
dust-continuum-identified MC clumps \citep{2013ApJ...770...39E} is therefore
consistent with that of the 77~km~s$^{-1}$ MCs.


The mean column density from C$^{18}$O~($J$=1--0) emission is 
0.5--1.5$\times10^{23}$~cm$^{-2}$ for the regions of FP1, FP3, FP4, and FP5, 
in which regions the column density from NH$_{3}$ 
emission are all higher than 1$\times10^{23}$cm$^{-2}$ \citep{2012A&A...544A.146W}. 
The FP1, FP3, FP4, and FP5 indeed show the densest regions along the filament
(see the red contours in the left panel of Figure~8), whereas the 
dense molecular gas in FP2 is exhausted and dissipated by the very strong star formation 
activities there.
The abundance ratio $X_{^{13}{\rm CO}}$/$X_{{\rm C}^{18}{\rm O}}$ is higher than 10 in the
region of FP2, which is consistent with the result that the abundance ratio becomes higher 
in the mature \HII\ region within the GMC. 

It is worth noting that the column density of FP1 derived from the C$^{18}$O
emission is comparable to that from the 870~$\mu$m dust emission.
According to \cite{2010ApJ...723L...7K}, both FP1 and FP5 are
potentially dense enough ($m_{\rm clump}>m_{\rm lim}[r]=870\Msun~[r/{\rm pc}]^{1.33}$)
to form massive stars. On the other hand, the 870~$\mu$m fluxes
of FP1 (3.68~Jy) and FP5 (4.93~Jy) are just close to the conservative criterion
of 5~Jy at 4.5~kpc, which is the flux limit at the distance to potentially
form high-mass stars \citep{2014A&A...565A..75C}.

Therefore, two dense clumps potentially sustain high-mass star 
formation (FP1 and FP5), two ongoing massive stars (FP3 and FP4),
and one mature \HII/SNR complex region (FP2), which are all the dense
gas concentrations associated with the filament, can well represent the fragmentation
on large scales ($\sim$10~pc) along the dense filamentary GMC G23.0$-$0.4.

\subsubsection{Turbulence Dominates the Large-scale Fragmentation}
We presented the main physical properties of the dense filament
GMC G23.0$-$0.4 in Section 3.1. The large-scale fragmentation of the dense
gas along the filament was discussed in Section 4.3.1. 
In the section, we investigate the stability of the dense filament.
 
If we regard the dense gas traced by the C$^{18}$O~($J$=1--0) emission
as a whole, the stability of the filamentary GMC can be described by the 
virial parameter $\alpha=M_{\rm vir}/M=2\sigma_{v}^{2}l/GM$
\citep{1992ApJ...395..140B,2000MNRAS.311...85F},
where $\sigma_{v}=\Delta V_{\rm C18O}/2.355$, $l$, and $G$ are 
the average velocity dispersion of C$^{18}$O~($J$=1--0), the 
length of the dense filament,
and the gravitational constant, respectively.
The virial parameter $\alpha$ thus is estimated to be 0.27, which indicates
that the dense filament is gravitationally bound.
Then the dense GMC is unstable and will collapse on a free-fall timescale if 
no other supporting mechanisms are included in the whole system.
In the above calculation, we take the long dense GMC as a symmetrica cylinder.
The parameters of the cylinder ($\sim66'\times6'\times6'$) can be obtained from 
the CO observations (see Section 3.1.1 and Table~2). 
The $\sim66'$ length of GMC G23.0$-$0.4 can be gotten from the \thCO\
emission ($\sim$1.2~K cutoff of \thCO\ emission, see the second contour level in
the middle panel of Figure~6). Assuming the length of 
the C$^{18}$O emission of $\sim66'$, the 
mean width of the dense filament is Area(C$^{18}$O)/length$\sim6'$.
The morphology of the dense filament also can be seen in Figures 4 and 8,
which clearly show the boundary of the GMC.
We also note that the virial parameter of the GMC from the \thCO\ emission is similar to 
that from the C$^{18}$O emission. It is normal since the value of the length, the
velocity dispersion, and the total mass of the GMC from \thCO\ and C$^{18}$O
are similar (see Table~2).

Moreover, we can estimate the fragmentational separation within the
large-scale filamentary structure to determine the character of
the gaseous cylinder.
Adopting 10$^{3}$~cm$^{-3}$ as the mean density of the GMC (Table~2),
the filament scale height $H$=$\sigma_{v}(4G\pi\rho_{c})^{-0.5}$ is about
0.8~pc due to the sausage instability of a self-gravitating fluid cylinder
\citep{1953ApJ...118..116C,1992ApJ...388..392I}. In the above calculation, we
use the mean full width at half-maximum (FWHM) of C$^{18}$O~($J$=1--0)
emission ($\Delta V_{\rm C18O}=$3.5~km~s$^{-1}$) in the GMC to estimate the
velocity dispersion of the turbulence \citep{2000MNRAS.311...85F}.
It leads to a spacing of $22H\sim$17~pc ($\sim$0\fdg22 at 4.4~kpc) between 
the fragmentation clumps, which result is in good agreement with the 
observations of the fragmentation spacing 0\fdg18--0\fdg26 along the 
dense filamentary GMC G23.0$-$0.4 (see Figure~9).
It indicates that the non-thermal turbulent pressure dominates 
over thermal pressure, which is consistent with other works
\citep[e.g.,][]{2010ApJ...719L.185J,2013A&A...557A.120K}.
In such case, the turbulence seems to control the 
fragmentation of the dense gaseous filament on large scale.

 
The regions of the five FPs represent the nodes of the filamentary GMC, in which
regions a large amount of molecular gas fragments into an assembly of cores 
of subsequent protostars 
\citep[see the sketch of Figure~2 in][]{2010ApJ...719L.185J}.  
If these cores are massive and dense enough to form high-mass stars, the stellar feedback
such as outflows, UV radiation field, stellar winds, and supernova explosions
within the dense filament
will dramatically affect the surrounding molecular gas. FPs 2--4 in GMC G23.0$-$0.4
is probably in such case. The mass in these regions is indeed massive enough 
to form massive stars. FP2, with its mature \HII/SNR complex region,
represents a slightly older generation than FP3 and FP4. The dense molecular gas in
the region of FP2 was now exhausted, whereas the dense molecular gas in regions of
FP3 and FP4 is still plentiful (Figures~8 and 9). 
Actually, recent massive star formation are ongoing in FP3 and FP4,
which are revealed by the 6.7~GHz methanol masers \citep{2002A&A...392..277S}.
FPs 1 and 5, which are at both ends of the dense filament,  
are also massive and dense enough to form massive stars (Section 4.3.1).

\section{SUMMARY}
We have presented a detail CO line study of the dense filamentary 
GMC G23.0$-$0.4. 
Combining the molecular line observations and the 870~$\mu$m dust continuum emission, the 
main results are summarized as follows:

1. GMC G23.0$-$0.4 with LSR velocity of 77~km~s$^{-1}$ displays filamentary 
structure roughly along the Galactic plane, which also shows hierarchical branch 
structure in C$^{18}$O~($J$=1--0) emission. Some slim filaments and pc-scale dense 
clumps are distributed along the main body of the GMC.

2, The optical depth of \thCO~($J$=1--0), $\tau_{13}$, is higher than 0.5 in the 
dense parts of the giant molecular filament, while $\tau_{18}$ is only close to 
0.2 in the same regions. It indicates that C$^{18}$O is indeed optically thin in the 
dense regions of the GMC. The dense parts of the GMC is readily seen in
C$^{18}$O~($J$=1--0) emission, which is also the good tracer of the high-density 
molecular gas ($\gsim5\times10^{22}$~cm$^{-3}$ in our case, see Figure~9).

3, The mass of the dense part of GMC G23.0$-$0.4 is $\sim5\times10^{5}~\Msun$ in the 
region of $\sim84\times$15~pc, while the total mass of the GMC is probably
close to $1.2\times10^{6}~\Msun$. 

4, GMC G23.0$-$0.4, which has a distance of 4.4~kpc, is probably 
located at the near side of the Scutum-Centaurus arm in the inner Galaxy. 
The long structure at $b\sim-$0\fdg4, which is traced by C$^{18}$O~($J$=1--0) 
emission along the longitude, is probably related to the density wave of 
a spiral arm of the Milky Way.

5, There are at least two SNRs, several \HII\ regions, and many massive stars 
associated with the 77~km~s$^{-1}$ molecular gas, indicating massive star formation 
occurred several Myr ago in the region. The natal GMCs of these objects probably have 
been destroyed by the strong stellar feedback. Meanwhile, these energetic sources are 
interacting with the nearby GMC G23.0$-$0.4.

6, The distribution of the abundance ratio ($X_{^{13}{\rm CO}}$/$X_{{\rm C}^{18}{\rm O}}$) 
is coincident well with the distribution of the cold dust emission along the dense filamentary 
GMC G23.0$-$0.4. The giant filament is massive and dense enough to
form high-mass stars.
Actually, massive star formation is ongoing in the nodes of the giant filament,
which is also traced by the bright dust-continuum emission, the 6.7~GHz methanol 
maser, and \HII$+$\HII/SNR complex associated with the dense GMC. The fragmentation 
with spacing $\sim$0\fdg22 along the filamentary GMC G23.0$-$0.4 also can be explained 
by the sausage instability of the cylinder.

7, The turbulence seems to control the
fragmentation process of the dense gaseous filament on large scale.
Combing the millimeter and IR dataset, we show that massive star
formation is ongoing at the fragmentation points with the periodic spacing along 
the entire filament. The massive stellar feedback near the dense filament 
has little influence on the evolution of the filamentary GMC G23.0$-$0.4.
On the contrary, the evolutionary processes of the massive stars within 
the massive filament dominate the fate of the dense GMC.

\acknowledgments
The authors acknowledge the staff members of the Qinghai Radio
Observing Station at Delingha for their support of the observations.
We would like to thank the anonymous referee for the critical 
comments and suggestions which helped to improve the paper.
This work is supported by NSFC grants 11233001 and 11233007.
The work is a part of the Multi-Line Galactic Plane Survey in CO and its
Isotopic Transitions, also called the Milky Way Imaging Scroll
Painting, which is supported by the Strategic Priority Research Program,
the Emergence of Cosmological Structures of the Chinese Academy of 
Sciences, grant No. XDB09000000.

\bibliographystyle{apj}
\bibliography{references}

\begin{deluxetable}{cccccc}
\tabletypesize{\scriptsize}
\tablecaption{Details of the Survey Data Used in the Paper}
\tablehead{
\colhead{\begin{tabular}{c}
Name \\
   \\
\end{tabular}} &
\colhead{\begin{tabular}{c}
Survey Range         \\
  \\
\end{tabular}} &
\colhead{\begin{tabular}{c}
Line/Continuum\\
  \\
\end{tabular}} &
\colhead{\begin{tabular}{c}
Resolution      \\
 \\
\end{tabular}} &
\colhead{\begin{tabular}{c}
rms      \\
 \\
\end{tabular}} &
\colhead{\begin{tabular}{c}
Reference      \\
 \\
\end{tabular}}
}
\startdata
COHRS    & 10$^{\circ}<l<65^{\circ}$     & \twCO~($J$=3--2)  & 16$''$, 1~km~s$^{-1}$  & $\sim$1~K & 1 \\
         &  -0\fdg25$\leq b \leq$0\fdg25   &                   &                        &     &   \\
\hline
GRS      & 18$^{\circ}<l<$55\fdg7        & \thCO~($J$=1--0)  & 46$''$, 0.2~km~s$^{-1}$  & $\sim$0.3~K & 2 \\
         &  -1$^{\circ}<b<1^{\circ}$     &                   &                        &     &   \\
\hline
VGPS     & 18$^{\circ}<l<67^{\circ}$     &  \mbox{H\,\textsc{i}} line & 1$'$, 1.56~km~s$^{-1}$  & $\sim$2~K & 3 \\
         &  -1\fdg2$<l<$1\fdg2           &  21~cm continuum      & 1$'$                    & $\sim$0.3~K  &   \\
\hline
ATLASGAL & -80$^{\circ}<l<60^{\circ}$    &  870~$\mu$m continuum  & 16\farcs7   & $\sim$50-70mJy/beam & 4,5,6 \\
         &  -1$^{\circ}<b<1^{\circ}$     &                   &                        &     &   \\
\enddata
\tablecomments{
(1)~\citealp{2013ApJS..209....8D}; (2)~\citealp{2006ApJS..163..145J};
(3)~\citealp{2006AJ....132.1158S}; (4)~\citealp{2009A&A...504..415S};
(5)~\citealp{2013A&A...549A..45C}; (6)~\citealp{2014A&A...565A..75C}.
}
\end{deluxetable}

\begin{deluxetable}{ccccc}
\tabletypesize{\scriptsize}
\tablecaption{Properties of GMC G23.0$-$0.4 }
\tablehead{
\colhead{\begin{tabular}{c}
Molecule $^{\mathrm {a}}$ \\
tracer     \\
\end{tabular}} &
\colhead{\begin{tabular}{c}
Area          \\
(arcmin$^{2}$)  \\
\end{tabular}} &
\colhead{\begin{tabular}{c}
$N$(H$_2$) \\
(10$^{22}$~cm$^{-2}$)\\
\end{tabular}} &
\colhead{\begin{tabular}{c}
$M$(H$_2$) $^{\mathrm {b}}$      \\
($10^{5}~\Msun$)\\
\end{tabular}} &
\colhead{\begin{tabular}{c}
$n$(H$_2$) $^{\mathrm {b,c}}$      \\
(cm$^{-3}$)\\
\end{tabular}}
}
\startdata
\twCO~($J$=1--0) & 877 & 2.1 & 5.2$\du^2$ & 890$\du^{-1}$ \\
\thCO~($J$=1--0) & 863 & 1.7 & 4.2$\du^2$ & 730$\du^{-1}$ \\
C$^{18}$O~($J$=1--0) & 388 & 2.9 & 3.2$\du^2$ & 1200$\du^{-1}$
\enddata
\tablecomments{
$^{a}$ See text for the methods used for the calculation.
$^{b}$ Parameter $\du$ is the distance to the GMC in units of 4.4~kpc.
$^{c}$ Assuming a depth of 6$'$, the mean width of the filament
in C$^{18}$O~($J$=1--0) emission, for GMC G23.0$-$0.4 along the LOS.}
\end{deluxetable}

\begin{deluxetable}{cccccc}
\tabletypesize{\scriptsize}
\tablecaption{Characters of the fragmentation nodes along GMC G23.0$-$0.4 }
\tablehead{
\colhead{\begin{tabular}{c}
Name \\
\end{tabular}} &
\colhead{\begin{tabular}{c}
($l$, $b$)    \\
\end{tabular}} &
\colhead{\begin{tabular}{c}
Tracer $^{\mathrm {a}}$    \\
\end{tabular}} &
\colhead{\begin{tabular}{c}
$T_{\rm ex}$ $^{\mathrm {b,c}}$ \\
\end{tabular}} &
\colhead{\begin{tabular}{c}
$\frac{X_{^{13}{\rm CO}}}{X_{{\rm C}^{18}{\rm O}}}$ $^{\mathrm {c}}$     \\
\end{tabular}} &
\colhead{\begin{tabular}{c}
Evolutionary stage    \\
\end{tabular}}
}
\startdata
FP1          & (22\fdg55, $-$0\fdg52)   & 
bright dust-continuum emission (3.76~Jy at 870~$\mu$m) $^{\mathrm {1,2}}$  & 
$\sim$15~K   & $\sim$4--5              &  early stage    \\
             &                         & 
high mass clumps from NH$_{3}$ ($>1\times10^{23}$cm$^{-2}$) $^{\mathrm {3}}$     &
             &                         &  \\
\hline
FP2          & (22\fdg76, $-$0\fdg48)   & 
\HII/SNR complex $^{\mathrm {4,5,6,7}}$ &
$\sim$20~K   & $\sim10$                &  mature stage    \\
\hline
FP3          & (23\fdg01, $-$0\fdg41)   &
bright dust-continuum emission (12.76~Jy at 870~$\mu$m) $^{\mathrm {1,2}}$ &
$\sim$25~K   & $\sim4$     	       & ongoing massive  \\
             &                         &
high mass clumps from NH$_{3}$ ($>3\times10^{23}$cm$^{-2}$) $^{\mathrm {3}}$     &
             &                         &  star formation \\
    	     &                         & 
6.7~GHz methanol maser $^{\mathrm {8,9,10}}$                             &
             &                         &   \\
\hline
FP4 	     & (23\fdg20, $-$0\fdg38)  & 
bright dust-continuum emission (11.3~Jy at 870~$\mu$m) $^{\mathrm {1,2}}$  &
$\sim$20~K   & $\sim$4--5              & ongoing massive   \\
             &                         &
high mass clumps from NH$_{3}$ ($>3\times10^{23}$cm$^{-2}$) $^{\mathrm {3}}$     &
             &                         &  star formation \\
    	     &                         & 
6.7~GHz methanol maser $^{\mathrm {9,11}}$                               &
             &                         &    \\
\hline
FP5          & (23\fdg36, $-$0\fdg29)  &
bright dust-continuum emission (4.93~Jy at 870~$\mu$m) $^{\mathrm {1,2}}$  &
$\sim$15~K   & $\sim$5--7              &  early stage    \\
             &                         &
high mass clumps from NH$_{3}$ ($>1\times10^{23}$cm$^{-2}$) $^{\mathrm {3}}$     &
             &                         &   \\
\enddata
\tablecomments{
$^{a}$ (1)~\citealp{2010ApJS..188..123R}; (2)~\citealp{2014A&A...565A..75C};
(3)~\citealp{2012A&A...544A.146W};
(4)~\citealp{1997ApJ...488..224K}; (5)~\citealp{2006AJ....131.2525H};
(6)~\citealp{2011ApJS..194...32A}; (7)~\citealp{2014ApJS..212....1A};
(8)~\citealp{1995MNRAS.272...96C}; (9)~\citealp{2002A&A...392..277S};
(10)~\citealp{2009ApJ...693..424B}; (11)~\citealp{2000A&AS..143..269S}.
$^{b}$ Estimated from the peak temperature of \twCO~($J$=1--0).
$^{c}$ This work.
}
\end{deluxetable}

\begin{figure*}
\includegraphics[trim=0mm 0mm 0mm 150mm,scale=0.5,angle=270]{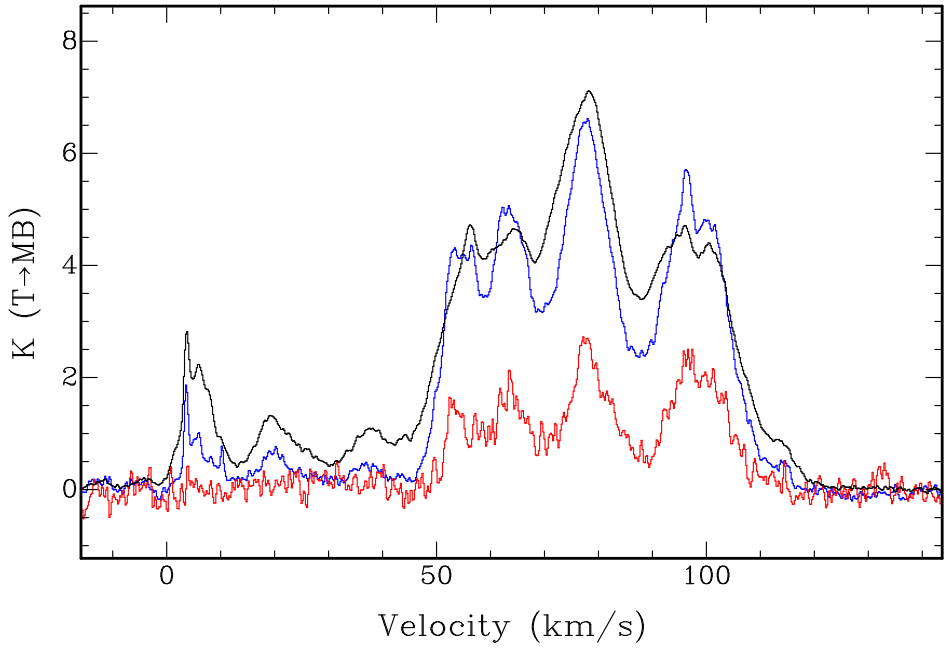}
\caption{\twCO\ ($J$=1--0; black), \thCO\ ($J$=1--0; blue, multiplied by
a factor of four), and C$^{18}$O ($J$=1--0; red, multiplied by
a factor of ten) spectra of the 30$'\times30'$ region.
\label{fig1}}
\end{figure*}

\begin{figure*}
\includegraphics[trim=0mm 0mm 0mm 150mm,scale=0.8,angle=0]{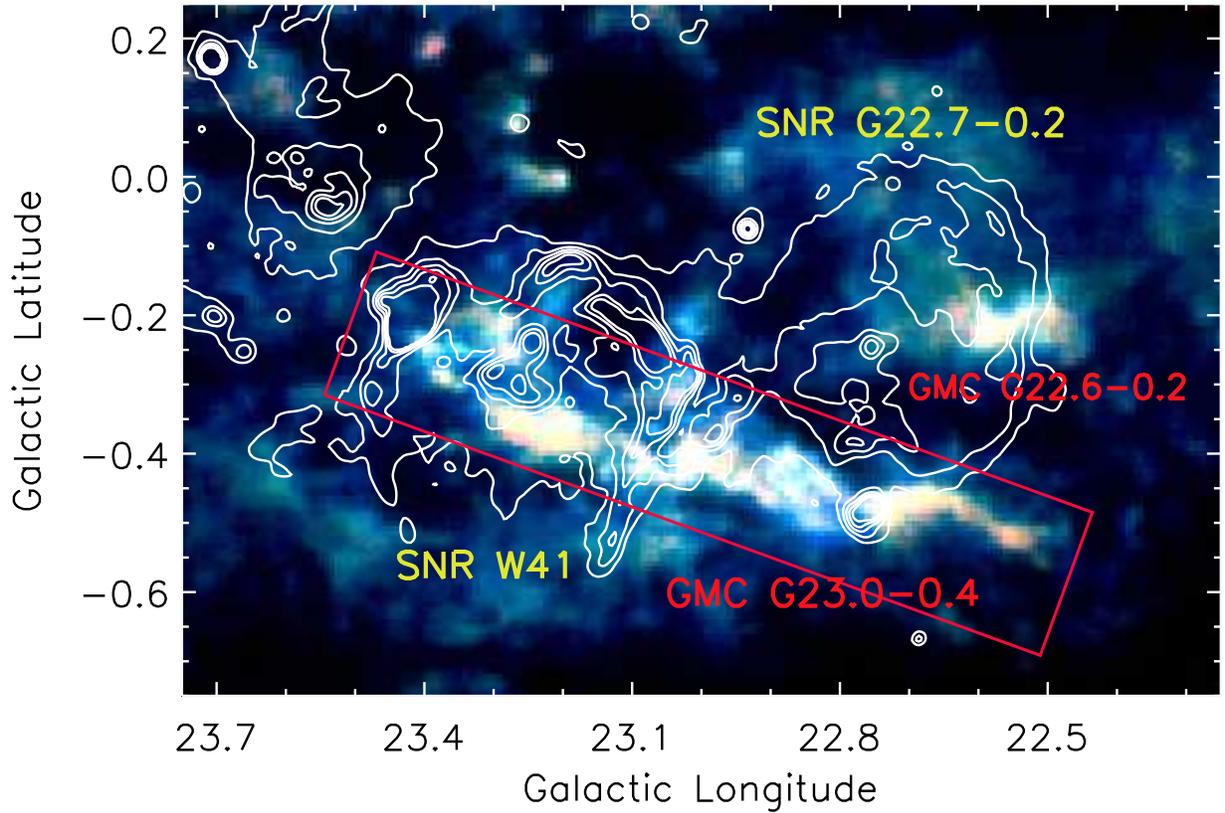}
\caption{\twCO\ ($J$=1--0; blue), \thCO\ ($J$=1--0; green), and
C$^{18}$O ($J$=1--0; red) intensity maps in the 74--79~km~s$^{-1}$
interval with a linear scale in the 1.5 square degree region
overlaid with the VGPS 1.4~GHz radio continuum emission contours.
The red rectangle shows the region of GMC G23.0$-$0.4.
SNR W41 (G23.3$-$0.3) and SNR G22.7$-$0.2 associated with the GMC
\citep{2014ApJ...796..122S} are labeled.
\label{fig2}}
\end{figure*}

\begin{figure*}
\includegraphics[trim=0mm 0mm 0mm 160mm,scale=0.45,angle=0]{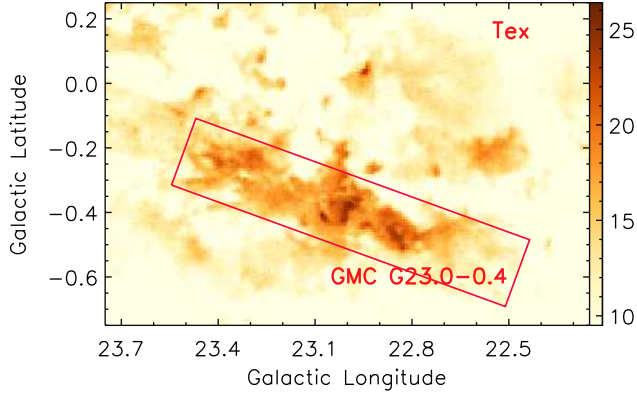}
\caption{Map of the excitation temperature with a linear scale in units of K from the 
optically thick \twCO\ ($J$=1--0) emission in the 74--79~km~s$^{-1}$ interval. The
red rectangle shows the region of GMC G23.0$-$0.4.
\label{fig3}}
\end{figure*}

\begin{figure*}
\centerline{
\includegraphics[trim=-15mm 0mm 0mm 160mm,scale=0.45,angle=0]{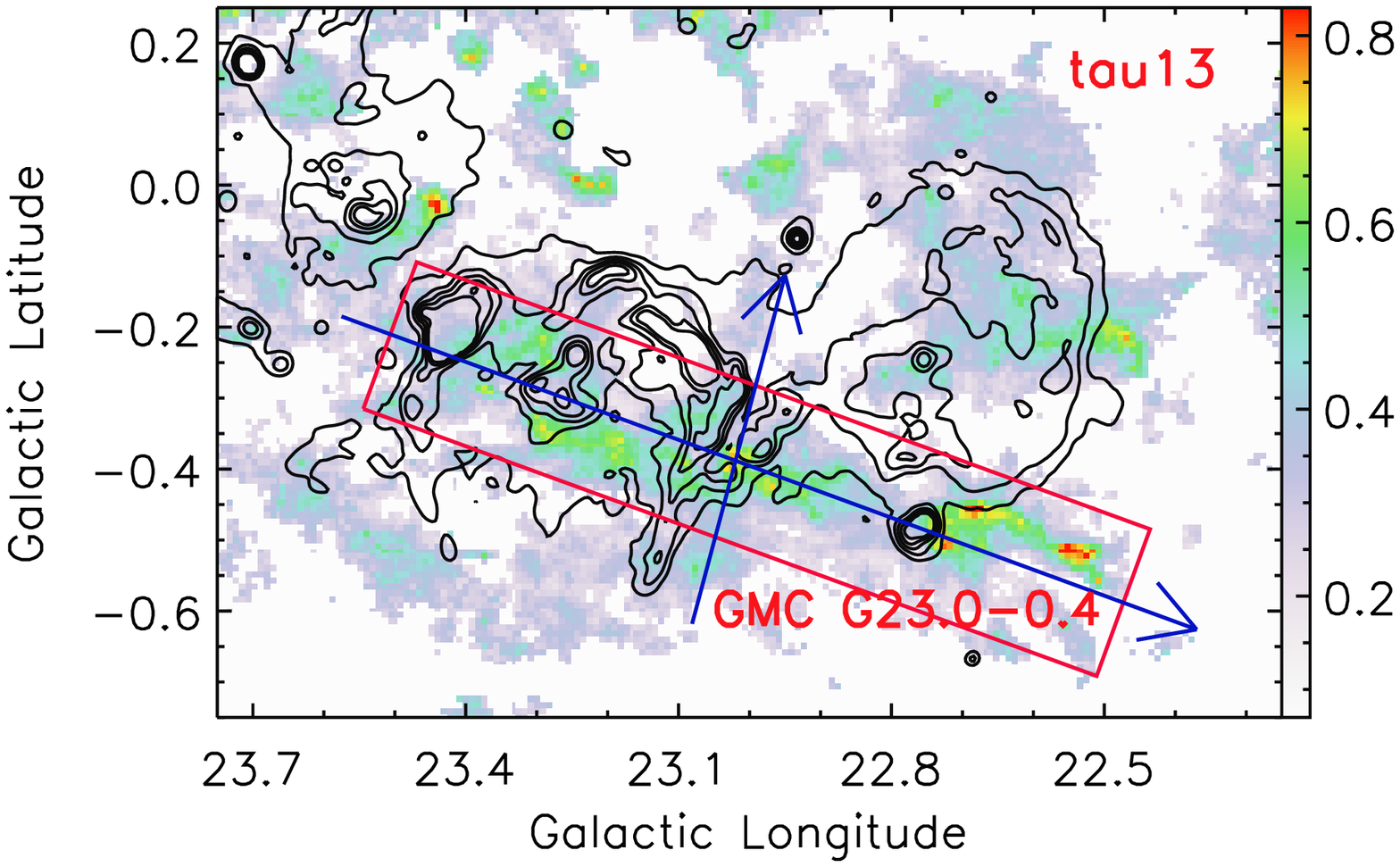}
\includegraphics[trim=20mm 0mm 0mm 160mm,scale=0.45,angle=0]{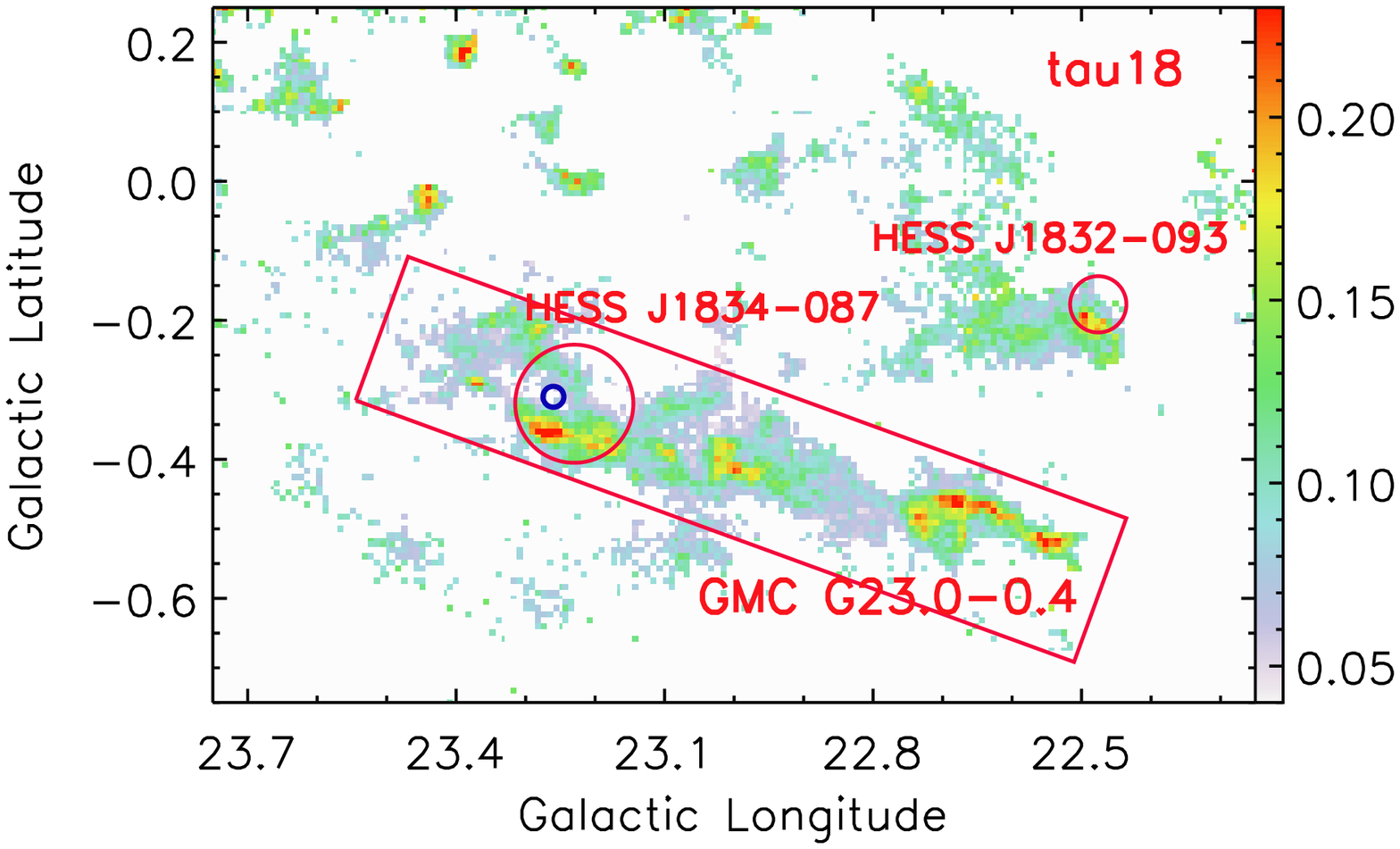}}
\caption{Maps of the optical depths of the \thCO\ ($J$=1--0) (left, with linear 
scale) and C$^{18}$O ($J$=1--0) (right, with square root scale). Note that the 
intensity scale of right image is different from the left one to highlight
the GMC structures. The blue arrows in left panel show the direction of the 
PV diagrams (see Figures~6,7).
The red circles in the right panel show the locations of 
HESS J1834--087 \citep{2015A&A...574A..27H} and 
HESS J1832--093 \citep{2015MNRAS.446.1163H}, respectively. The small blue 
circle indicates the location of 1720~MHz OH maser \citep{2013ApJ...773L..19F}.
\label{fig4}}
\end{figure*}

\begin{figure*}
\includegraphics[trim=0mm 0mm 0mm 30mm,scale=0.8,angle=0]{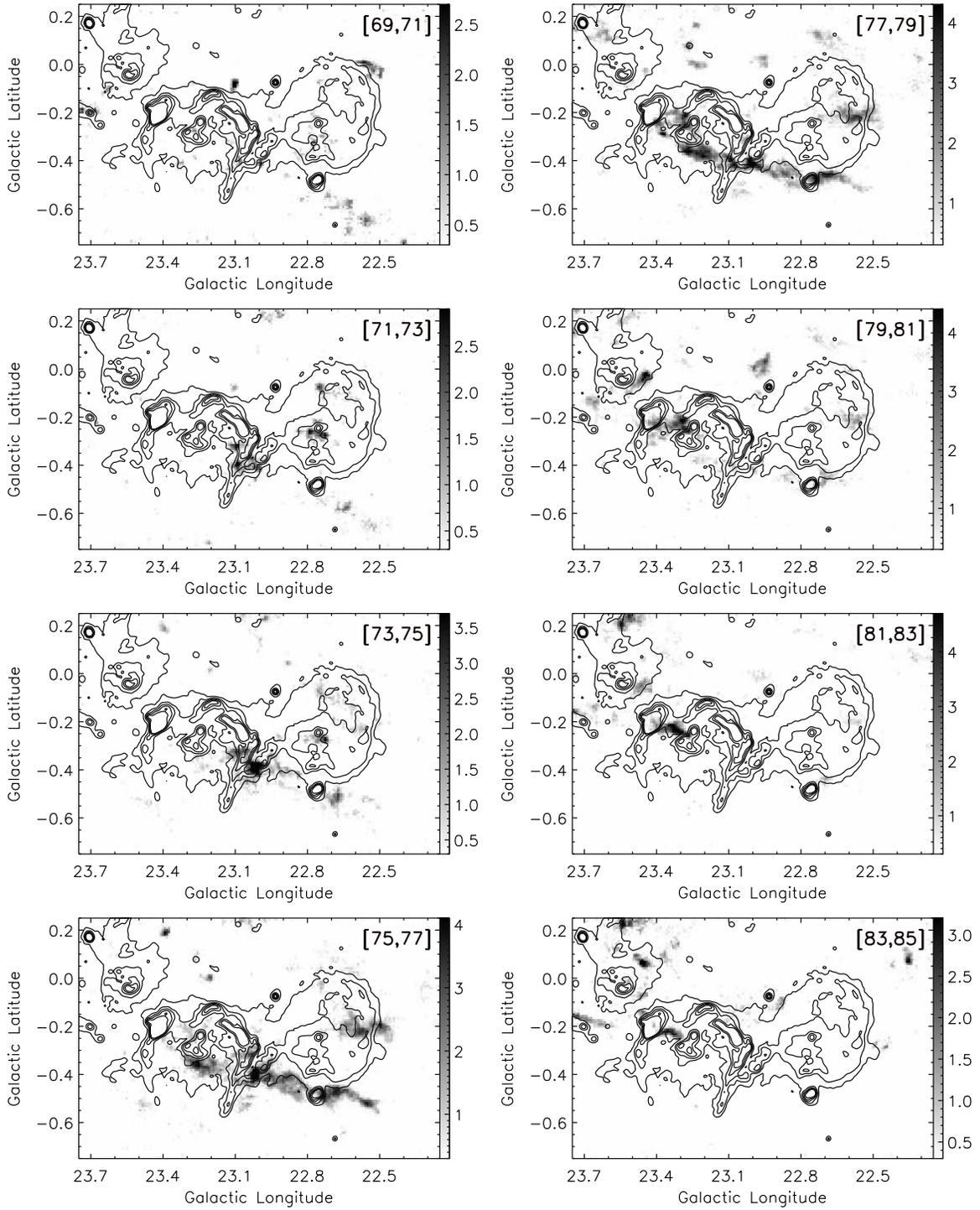}
\caption{
Velocity channel maps of the C$^{18}$O ($J$=1--0) emission line in units
of K~km~s$^{-1}$, overlaid with the VGPS 1.4~GHz radio continuum emission contours. 
The velocity range used for the integration is indicated
in the top-right conner of each panel. Note that the intensity scales
are different from each other.
\label{fig5}}
\end{figure*}

\begin{figure*}
\centerline{
\includegraphics[trim=0mm 0mm 0mm 20mm,scale=0.3,angle=0]{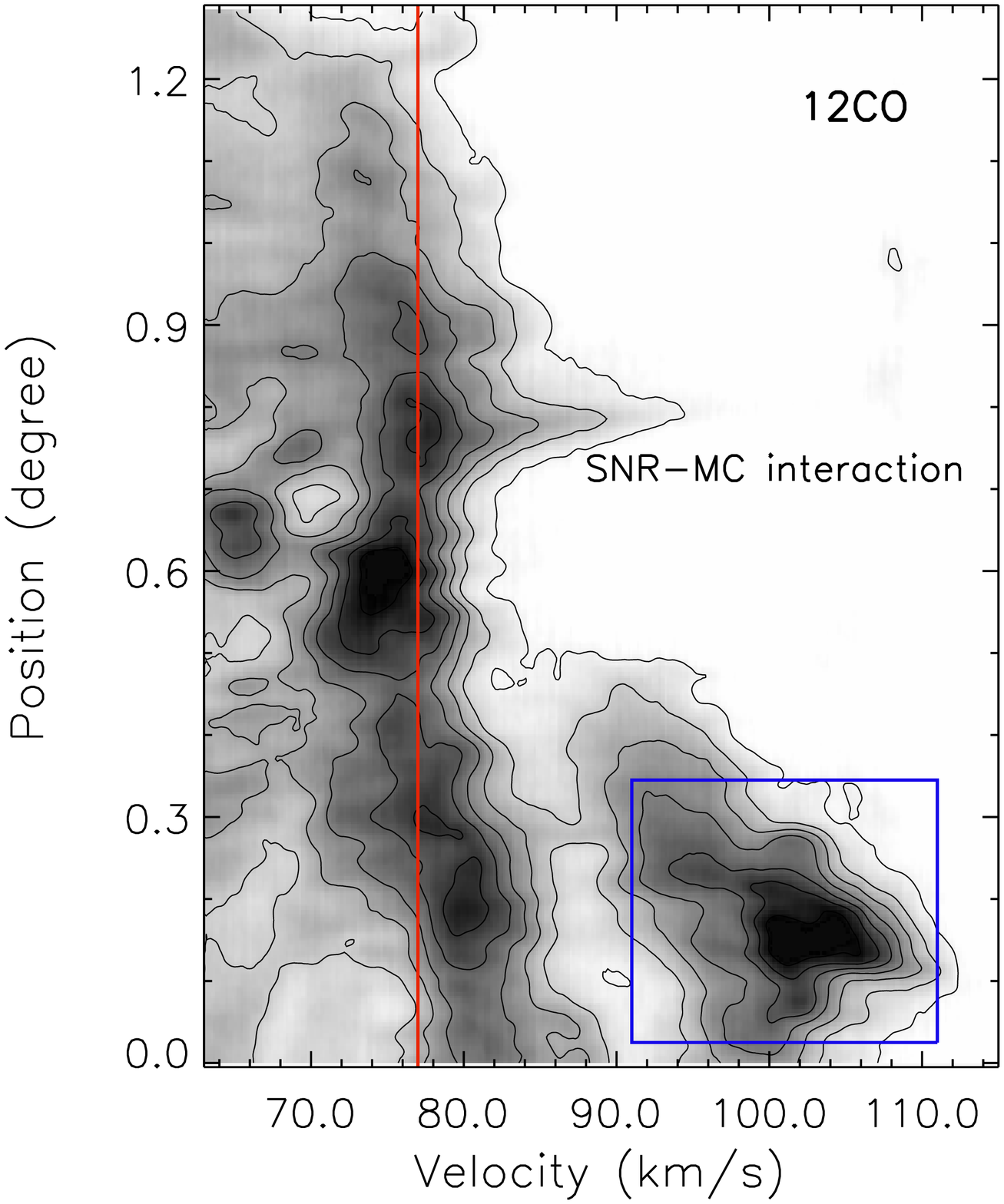}
\includegraphics[trim=0mm 0mm 0mm 20mm,scale=0.3,angle=0]{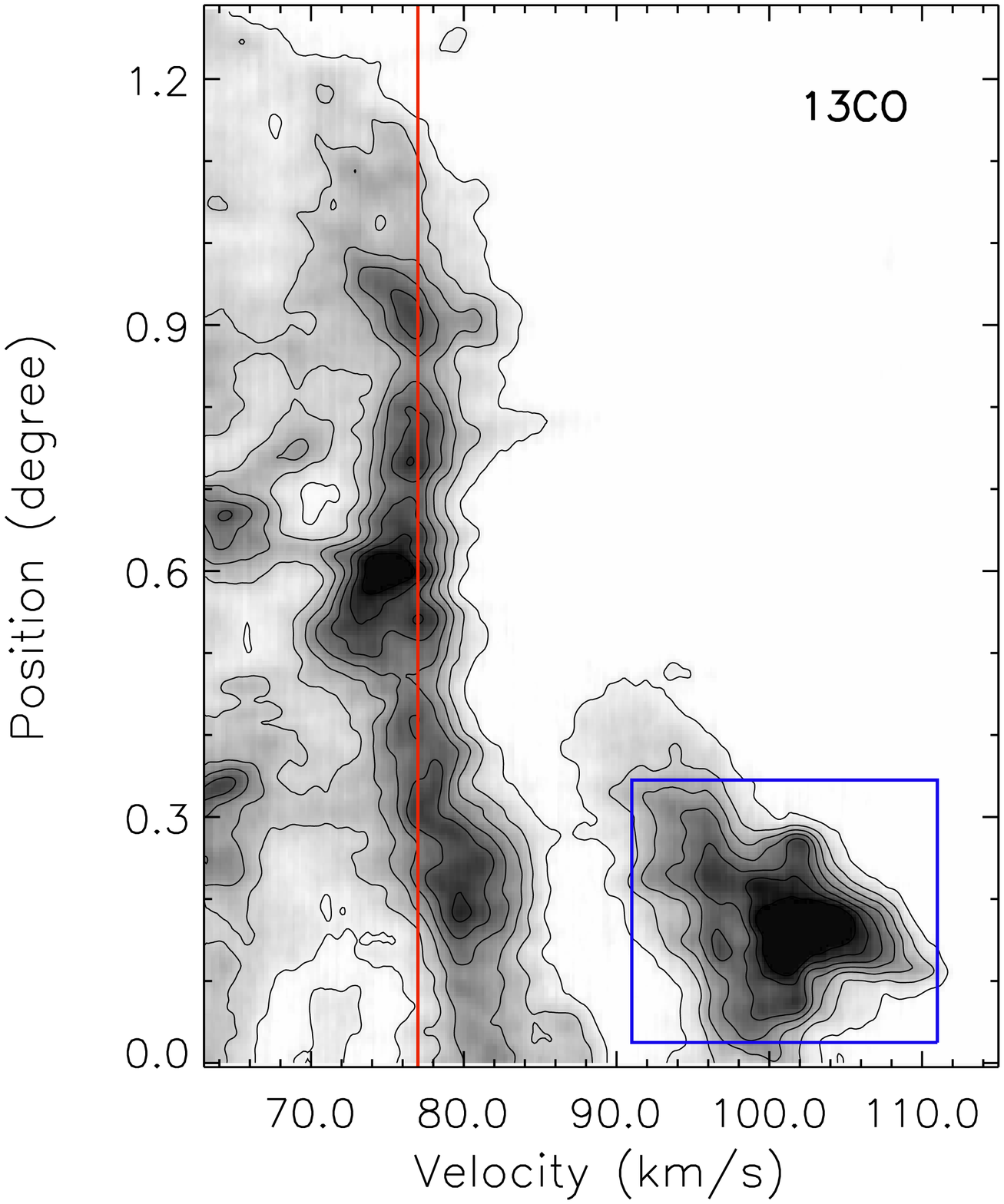}
\includegraphics[trim=0mm 0mm 0mm 20mm,scale=0.3,angle=0]{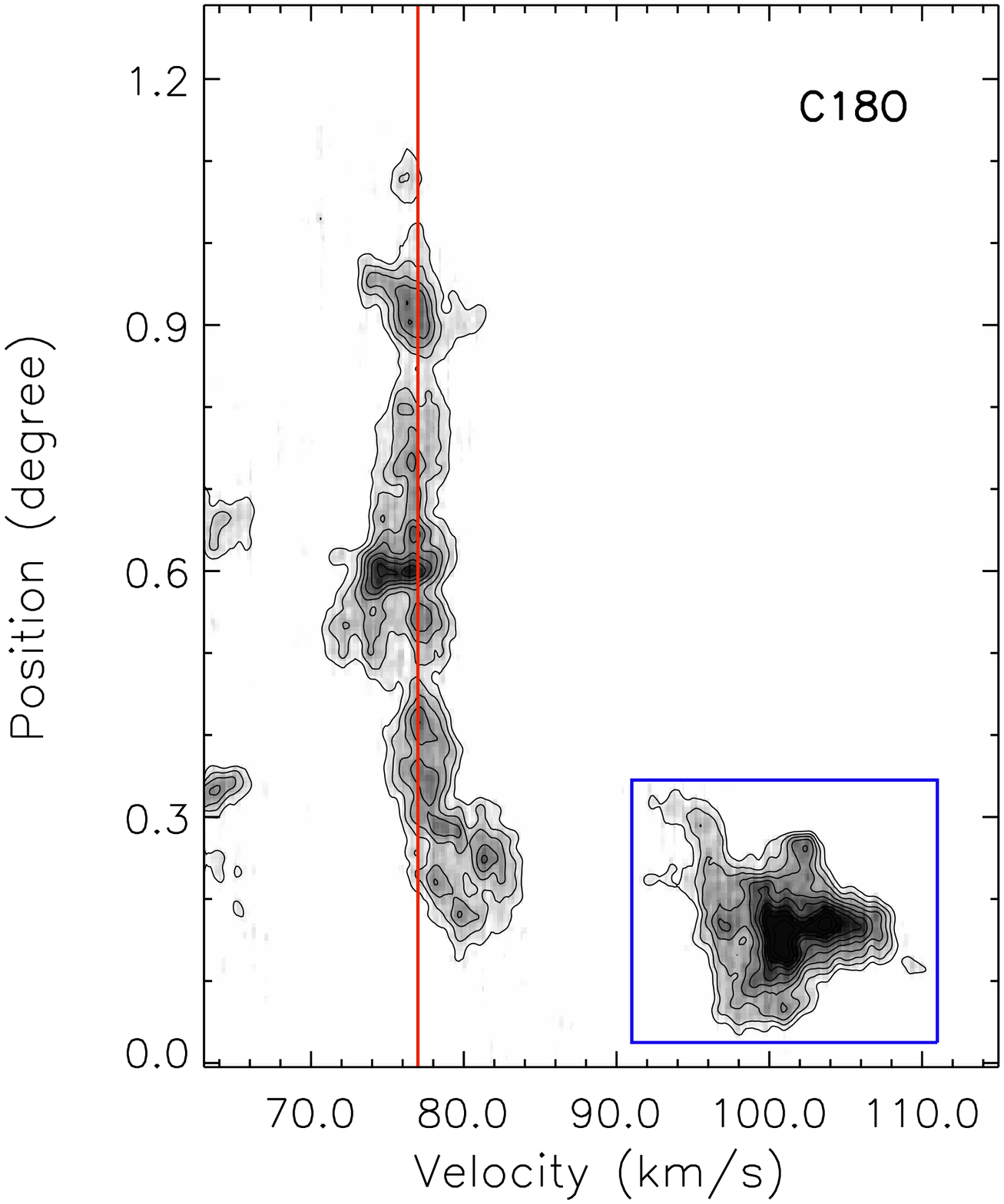}}
\caption{
PV diagrams of the \twCO\ ($J$=1--0), \thCO\ ($J$=1--0), and
C$^{18}$O ($J$=1--0) emission along the filamentary GMC G23.0$-$0.4 (see 
the long blue arrow in Figure~4). The position is measured along the long arrow
(($l$=23\fdg575, $b$=$-$0\fdg185) to ($l$=22\fdg370, $b$=$-$0\fdg626))
with a width of 10\farcm5.
The red line marks the LSR velocity of 77~km~s$^{-1}$. 
The blue rectangle shows the G23.4 cloud near the tangent point 
\citep{2012PASJ...64...74O}.
\label{fig6}}
\end{figure*}

\begin{figure*}
\includegraphics[trim=0mm 0mm 0mm 10mm,scale=0.4,angle=0]{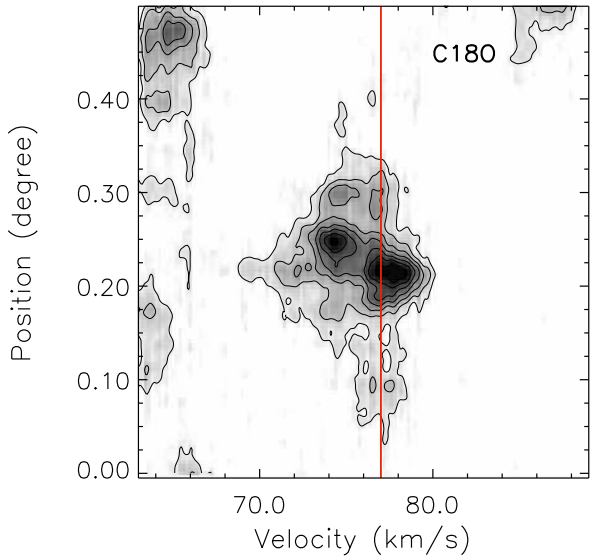}
\caption{
PV diagram of the 
C$^{18}$O ($J$=1--0) emission along the interface between SNR W41 and
SNR G22.7$-$0.2 (see the short blue arrow in Figure~4). The position is 
measured along the short arrow
(($l$=23\fdg081, $b$=$-$0\fdg618) to ($l$=22\fdg950, $b$=$-$0\fdg127))
with a width of 9\farcm5.
The red line marks the LSR velocity of 77~km~s$^{-1}$.
\label{fig7}}
\end{figure*}

\begin{figure*}
\centerline{
\includegraphics[trim=-10mm 0mm 0mm 190mm,scale=0.44,angle=0]{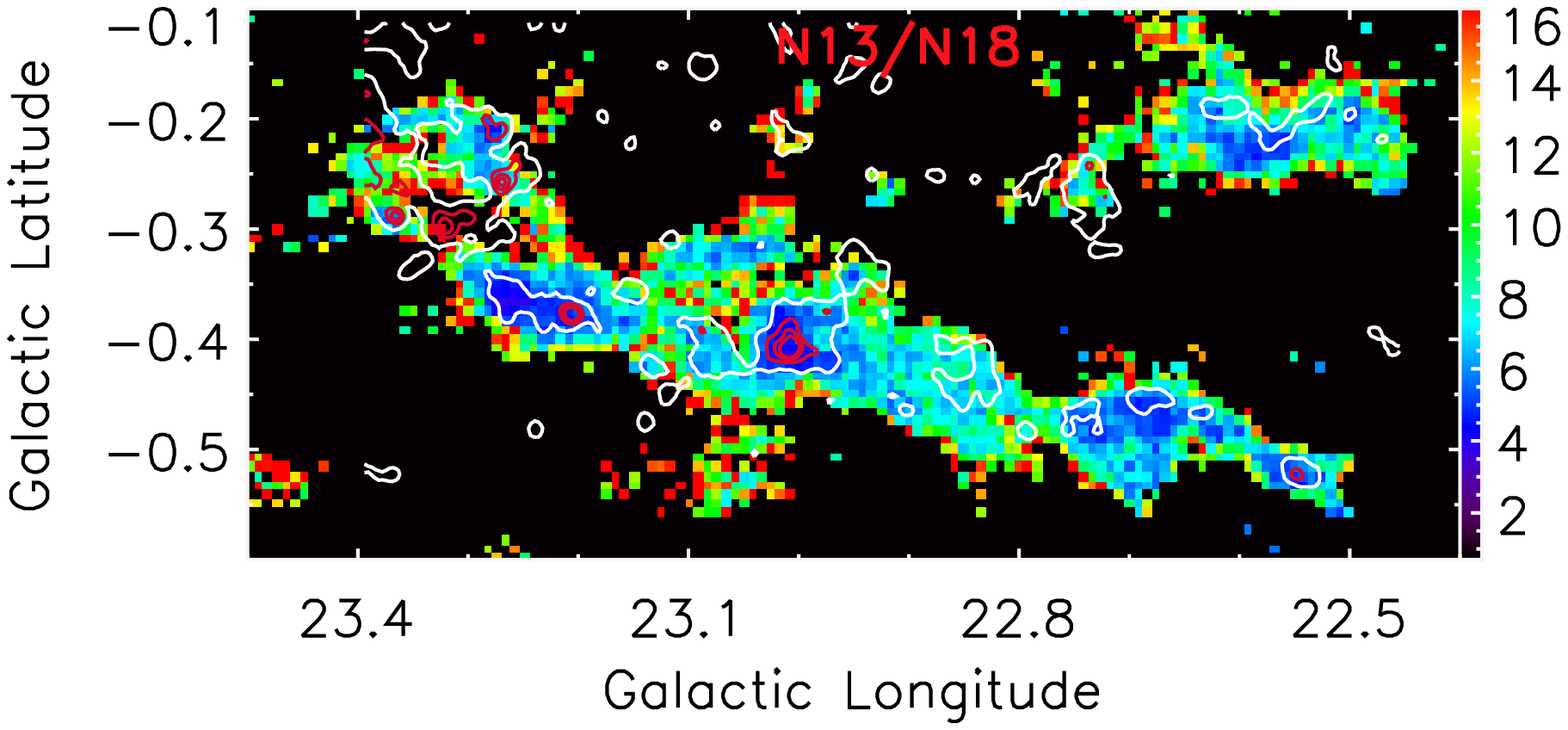}
\includegraphics[trim=20mm 0mm 0mm 190mm,scale=0.44,angle=0]{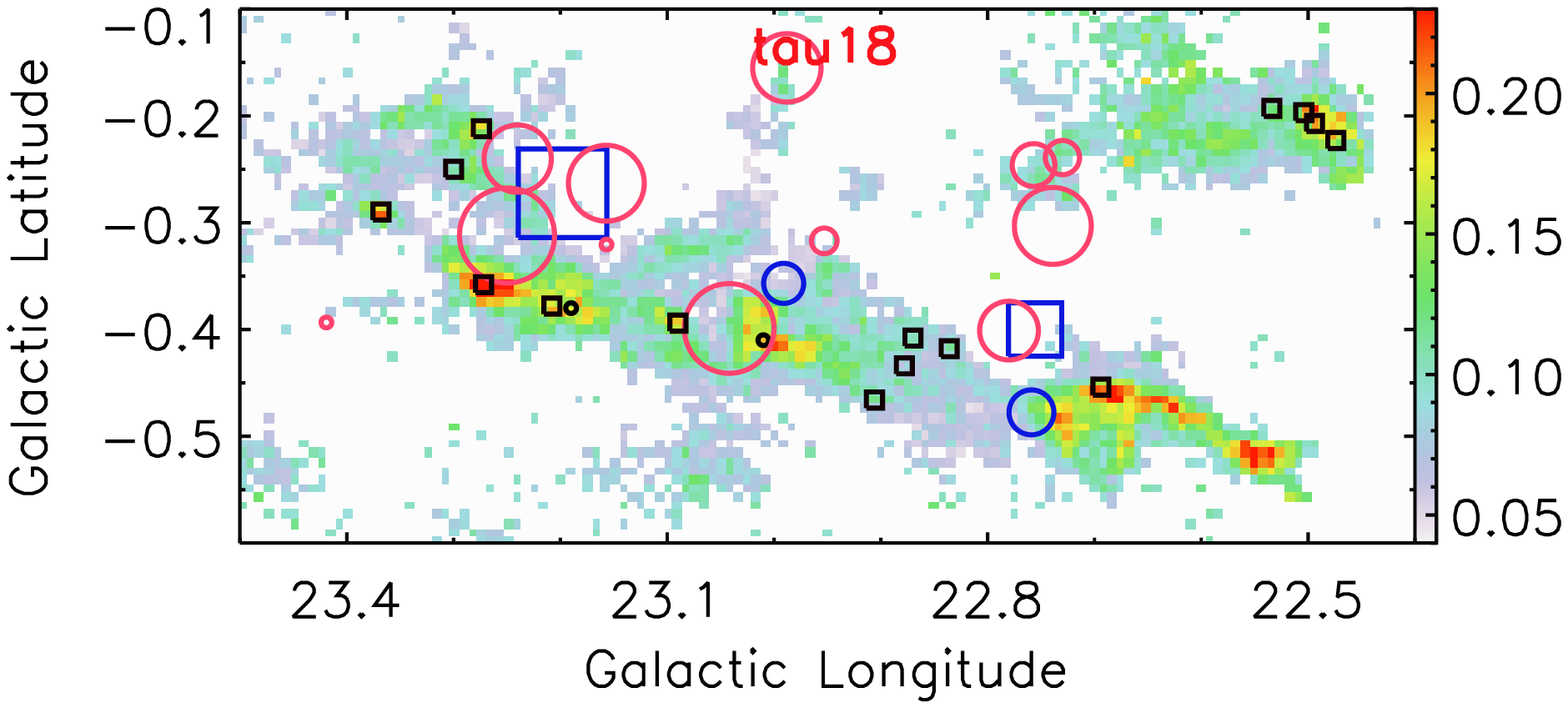}}
\caption{
Left panel: Map of the abundance ratio X$_{^{13}{\rm CO}}$/X$_{\rm C^{18}O}$ overlaid
with the ATLASGAL 870~$\mu$m continuum emission contours (0.2, 1.2, 2.2, and 3.2~Jy/beam).
The dense dust clumps ($>$1.2~Jy/beam) are highlighted in red contours. 
Right panel: Map of the optical depth of the C$^{18}$O ($J$=1--0) emission. 
The black boxes, black circles, red circles, blue circles, and blue boxes
show the fifteen dust-continuum-identified MC clumps, the two 6.7~GHz methanol 
masers, the twelve \HII\ regions, the two \HII/SNR complexes, and the two massive 
star groups, respectively.
\label{fig8}}
\end{figure*}

\begin{figure*}
\centerline{
\includegraphics[trim=-20mm 0mm 0mm 190mm,scale=0.8,angle=0]{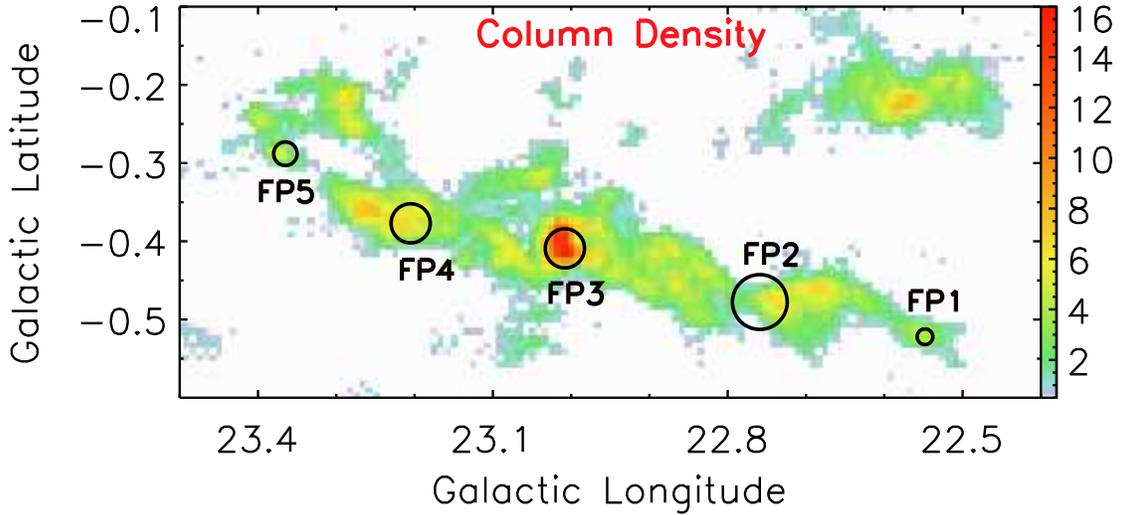}}
\caption{
Map of the column density from C$^{18}$O~($J$=1--0) emission with a logarithmic scale 
in units of 10$^{22}$~cm$^{-2}$. The four fragmentation points are marked with black circles.
\label{fig9}}
\end{figure*}

\end{document}